\documentclass[journal,twocolumn]{IEEEtran}

\usepackage{pgf,tikz}
\usetikzlibrary{arrows}
\usepackage{research5IEEE}

\usepackage{dsfont}
\usepackage{accents}
\usepackage{pifont}

\usepackage{amsmath}
\usepackage{color}
\usepackage{bbm}
\allowdisplaybreaks
\usepackage{cite}
\usepackage{url,epstopdf}
\usepackage{todonotes}
\newtheorem{theorem}{Theorem}
\newtheorem{definition}{Definition}
\newtheorem{lemma}{Lemma}
\newtheorem{proposition}{Proposition}
\newtheorem{corollary}{Corollary}
\newtheorem{remark}{Remark}
\newtheorem{example}{Example}

\newcommand\munderbar[1]{%
	\underaccent{\bar}{#1}}
\ifCLASSINFOpdf
\else
\fi

\title{Distributed Hypothesis Testing with Privacy Constraints}
\author{Selma Belhadj Amor,~\IEEEmembership{Member,~IEEE,}
	John~Doe,~\IEEEmembership{Fellow,~OSA,}
	and~Jane~Doe,~\IEEEmembership{Life~Fellow,~IEEE}
	\thanks{M. Shell was with the Department
		of Electrical and Computer Engineering, Georgia Institute of Technology, Atlanta,
		GA, 30332 USA e-mail: (see http://www.michaelshell.org/contact.html).}
	\thanks{J. Doe and J. Doe are with Anonymous University.}
	\thanks{Manuscript received April 19, 2005; revised August 26, 2015.}}

\author{Atefeh Gilani  $\quad$ Selma Belhadj Amor $\quad$  Sadaf Salehkalaibar     $\quad$  
        Vincent Y.~F.~Tan  
\thanks{A.~Gilani and S.~Salehkalaibar are with the  Electrical and Computer
Engineering Department, College of Engineering, University of Tehran (e-mail: \{atefehgilani,s.saleh\}@ut.ac.ir). S.~Belhadj Amor and V.~Y.~F.~Tan are with the  Department of Electrical and Computer
Engineering, National University of Singapore (e-mail:
\{elesba,vtan\}@nus.edu.sg).}  }

\begin{document}
\maketitle
%
%
\begin{abstract}
	We revisit the distributed hypothesis testing (or hypothesis testing with communication constraints) problem from the viewpoint of privacy. Instead of observing the raw data directly, the transmitter observes a sanitized or randomized version of it. We impose an upper bound on the mutual information between the raw and randomized data. Under this scenario, the receiver, which is also provided with side information, is required to make a decision on whether the null or alternative hypothesis is in effect. We first provide a general lower bound on the type-II exponent for an arbitrary pair of hypotheses. Next, we show that if the distribution under the alternative hypothesis  is the product of the marginals of the distribution under the null (i.e., testing against independence), then the exponent is known exactly. Moreover, we show that the strong converse property holds. Using ideas from Euclidean information theory, we also provide an approximate expression for the exponent when the communication rate is low and the privacy level is high. Finally, we illustrate our results with a binary and a  Gaussian example. 
\end{abstract}

\begin{IEEEkeywords}
Hypothesis testing, Privacy, Mutual information, Testing against independence, Zero-rate communication
\end{IEEEkeywords}

%
\IEEEpeerreviewmaketitle

\section{Introduction}

In the distributed hypothesis testing (or hypothesis testing with communication constraints) problem, some observations from the environment are collected by the sensors in a network. They describe these observations over the network which are finally received by the decision center. The goal is to guess the joint distribution governing the observations at terminals. In particular, there are two possible hypotheses $\mathcal{H}=0$ or $\mathcal{H}=1$, where the joint distribution of the observations is specified under each of them. The performance of this system is characterized by two criteria: the type-I and the type-II error probabilities. The probability of deciding on $\mathcal{H}=1$ (resp.\ $\mathcal{H}=0$) when the original hypothesis is $\mathcal{H}=0$ (resp.\ $\mathcal{H}=1$) is referred to as the type-I error (type-II error) probability. It is desired that the type-II error probability exponentially goes to zero as the blocklength $n$ grows to infinity, under a constrained type-I error probability.

A special case of interest is testing against independence where the joint distribution under $\mathcal{H}=1$ is the product of the marginals under $\mathcal{H}=0$. The optimal exponent of type-II error probability for testing against independence is determined by Ahlswede and Csisz\'ar in~\cite{AhlswedeCsiszar}. Several extensions of this basic problem are studied for a multi-observer setup~\cite{Lai, Kim, TianChen, Wagner, Gunduz}, a multi-decision center setup \cite{Michele, Michele2} and a setup with security constraints~\cite{Piantanida}. The main idea of the achievable scheme in these works is typicality testing \cite{Han, SHA}. The sensor finds a jointly typical codeword with its observation and sends the corresponding bin index to the decision center. The final decision is declared based on typicality check of the received codeword with the observation at the center. 
\begin{figure}[t]
	\centerline{
		\begin{tikzpicture}[line cap=round,>=triangle 45,x=1.0cm,y=1.0cm,xscale=1.1,yscale=0.7,thick]
		\draw [->][thick](2.3,0.5)--(3.1,0.5);
		\node at (2.7,0.8){$\hat{X}^n$};	
		\node at (0.8,0.8){$X^n$};
		\draw [thick] (1.2,0.1) rectangle (2.3,0.9);
		\node  at (1.75,1.3){Observer};
		\node at (1.75,0.5){$P_{\hat{X}^n|X^n}$};	
		\draw [->][thick](0.5,0.5)--(1.2,0.5);
		\draw [thick] (3.1,0.1) rectangle (4.1,0.9);
		\node at (3.6,0.5){Tx};
		\draw [->][thick](4.1,0.5)--(6.5,0.5);
		\node at (5.25,0.8){Message $M$};
		\node at (5.25,0.2){Rate $R$};
		\draw [thick] (6.5,0.1) rectangle (7.5,0.9);
		\node at (7,0.5){Rx};
		\draw [thick][->] (7,1.9)--(7,0.9);
		\node at (7.3,1.6){$Y^n$};
		\draw [thick][->] (7.5,0.5)--(8,0.5);
		\node at (8.3,0.6){$\set{\hat{H}}$};
		\end{tikzpicture}
	}	
	\caption{Hypothesis testing with communication and privacy constraints\label{fig1}}
\end{figure}
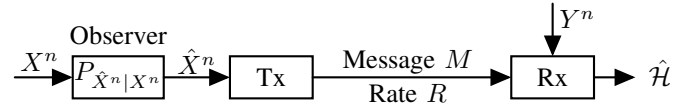
\subsection{Injecting Privacy Considerations Into our System}
We revisit the distributed hypothesis testing problem from a privacy perspective. In many applications such as healthcare systems, there is a need to randomize the data before publishing it. We use a privacy mechanism to sanitize the observation at the terminal before it is compressed; see Fig.~\ref{fig1}. The compression is performed at a separate terminal called \emph{transmitter}, which  communicates the randomized data over a noiseless link of rate $R$ to a receiver. The hypothesis testing is performed using the received data (the compression index and additional side information) to determine the correct hypothesis governing the original observations. The privacy criterion is defined by the mutual information \cite{Evfimievski,Smith,Sankar,liao18} of the published and original data.

There is a long history of research to provide appropriate metrics to measure privacy. 	To quantify the information leakage an observation $\hat{X}$ can induce on a latent variable $X$,
Shannon's mutual information $I(X;\hat{X})$ is considered in \cite{Evfimievski,Smith,Sankar,liao18}.  Smith \cite{Smith} proposed to use Arimoto's mutual information of order $\infty$, $I_\infty(X;\hat{X})$. 
Barthe and K\"opf \cite{Barthe-kopf,issa17, liao17} proposed the maximal information leakage $\max_{P_X} I_\infty(X;\hat{X})$. We refer the reader to \cite{Wagner-Eckhoff} for a survey on the existing information leakage measures. A different line of works, in statistics, computer science, and other related fields, concerns {\em differential privacy}, initially proposed in \cite{Dwork-DF}. Furthermore, a generalized notion---$(\epsilon,\delta)$-differential privacy  \cite{Dwork-epsilon-delta-DF}---provides a  unified mathematical framework for data privacy. The reader is referred to the survey by Dwork~\cite{Dwork-survey} and the statistical framework studied by Wasserman and Zhou~\cite{Wasserman-Zhou} and the references therein.

The privacy mechanism can be either memoryless or non-memoryless. In the former, the distribution of the randomized data at each time instant depends on the original sequence at the same time and not on the previous history of the data.

\subsection{Description of our System Model }
We propose a coding  scheme for the proposed setup. The idea is that the sensor, upon observing the source sequence, performs a typicality test and obtains its belief of the hypothesis.  If the belief is $\mathcal{H}=0$, it publishes the randomized data based on a specific memoryless mechanism. However, if its belief is $\mathcal{H}=1$,  it sends an all-zero sequence to let the transmitter know about its decision. The transmitter communicates the received data, which is a sanitized version of the original data or an all-zero sequence, over the noiseless link to the receiver. In this scheme, the whole privacy mechanism is non-memoryless since the typicality check of the source sequence which uses the history of the observation, determines the published data. It is shown that the achievable error exponent recovers previous results on hypothesis testing with zero and positive communication rates in \cite{Han}. 

A difference of the proposed scheme with some previous works is highlighted as follows.
The privacy mechanism even if it is memoryless, cannot be viewed as a noiseless link of a rate equivalent to the privacy criterion. Particularly, the proposed model is different from cascade hypothesis testing problem of \cite{Michele2} or similar works \cite{Kim, TianChen} which consider consecutive noiseless links for data compression and distributed hypothesis testing. The difference comes from the fact that in these works, a codeword is chosen jointly typical with the observed sequence at the terminal and its corresponding index is sent over the noiseless link. However, in our model, the randomized sequence is not necessarily jointly typical with the original sequence. Thus, there is a need for an achievable scheme which lets the transmitter know whether the original data is typical or not.

The problem of hypothesis testing against independence with a memoryless privacy mechanism is also considered. A coding  scheme is proposed where the sensor outputs the randomized data based on the memoryless privacy mechanism. The optimality of the achievable type-II error exponent is shown by providing a strong converse. Specializing the optimal error exponent to a binary example shows that an increase in the privacy criterion (a less stringent privacy mechanism) results in a larger type-II error exponent. Thus, there exists a trade-off between privacy and hypothesis testing criteria. The optimal type-II error exponent is further studied for the case of restricted privacy mechanism and zero-rate communication. The Euclidean approach of \cite{Borade, huang} is used to approximate the error exponent for this regime. The result confirms the trade-off between the privacy criterion and type-II error exponent. Finally, a Gaussian setup is proposed and its optimal error exponent is established.


\subsection{Main Contributions}

The contributions of the paper are listed in the following:
\begin{itemize}
	\item An achievable type-II error exponent is proposed using a non-memoryless privacy mechanism (Theorem~\ref{achievability_general} in Section~\ref{sec:general});
	\item The optimal error exponent of testing against independence with a memoryless privacy mechanism is determined. In addition, a  strong converse is also proved (Theorem~\ref{TAI_thm} in Section~\ref{sec:TAI});
	\item A binary example is proposed to show the trade-off between the privacy and error exponent (Section~\ref{ex:binary});
	\item A Euclidean approximation~\cite{Borade} of the error exponent is provided (Section~\ref{sec:Euc});
	\item A Gaussian setup is proposed and its optimal error exponent is derived (Proposition~\ref{Gaussian-prop} in Section~\ref{sec:Gaussian}).
	\end{itemize}

\subsection{Notation}

The notation mostly follows \cite{ElGamal}. Random variables are denoted by capital letters, e.g., $X$, $Y$, and their realizations by lower case lettes, e.g., $x$, $y$. The alphabet   of the random variable $X$ is denoted as $\mathcal{X}$. Sequences of random variables and their realizations are denoted by $(X_i,\ldots,X_j)$ and $(x_i,\ldots,x_j)$ and are abbreviated as $X_i^j$ and $x_i^j$. We use the alternative notation $X^j$ when $i=1$. Vectors and matrices are denoted by boldface letters, e.g., $\mathbf{k}$, $\mathbf{W}$. The $\ell_2$-norm of $\mathbf{k}$ is denoted as $\| \mathbf{k}\|$. The notation $\mathbf{k}^T$ denotes the transpose of~$\mathbf{k}$.

The probability mass function (pmf) of a discrete random variable $X$ is denoted as $P_X$, the conditional pmf of $X$ given $Y$ is denoted as $P_{X|Y}$. The notation $D(P_X\|Q_X)$ denotes the Kullback-Leibler (KL) divergence between two pmfs $P_X$ and $Q_X$. The total variation distance between two pmfs $P_X$ and $Q_X$ is denoted by $\big|  P_X-Q_X\big|=\frac{1}{2}\sum_x |P_X(x)-Q_X(x)|$.  We use $\text{tp}(x^n,y^n)$ to denote the joint type of $(x^n,y^n)$.  
 
For a given $P_{XY}$ and a positive number $\mu$, we denote by $\mathcal{T}_{\mu}^n(P_{XY})$, the set of jointly $\mu$-typical sequences~\cite{ElGamal}, i.e, the  set of all $(x^n,y^n)$ whose joint type is within $\mu$ of $P_{XY}$. The notation $\mathcal{T}^n(P_{X})$ denotes for the type class of the type $P_{X }$.

 The notation $h_{\text{b}}(\cdot)$ denotes the binary entropy function, $ h_{\text{b}}^{-1}(\cdot)$ its inverse over $\big[0,\frac{1}{2}\big]$, and $a\star b \eqdef a(1-b)+(1-a)b$ for $0\leq a,b\leq 1$. The differential entropy of a continuous random variable $X$  is $h(X)$. All  logarithms $\log(\cdot)$ are taken with respect to base $2$. 
 
 \subsection{Organization}
 
 The remainder of the paper is organized as follows. Section~\ref{sec:model} describes a mathematical setup for our proposed problem. Section~\ref{sec:general} discusses hypothesis testing with general distributions. The results for hypothesis testing against independence with a memoryless privacy mechanism are provided in Section~\ref{sec:positive-rate}. The paper is concluded in Section~\ref{sec:conclusion}.

\section{System Model}\label{sec:model}

Let $\set{X}$, $\set{Y}$, and $\hat{\set{X}}$ be arbitrary finite alphabets and let $n$ be a positive integer.
Consider the hypothesis testing problem with communication and privacy constraints depicted in Fig.~\ref{fig1}. The first terminal in the system, the \emph{Observer},  receives the sequence $X^n=(X_1,\ldots, X_n)\in \set{X}^n$ and outputs the sequence $\hat{X}^n=(\hat{X}_1, \ldots, \hat{X}_n)\in \hat{\set{X}}^n$, which is a noisy version of $X^n$ under a {\em privacy mechanism} determined by the conditional probability distribution $P_{\hat{X}^n|X^n}$; the second terminal, the \emph{Transmitter}, receives the sequence $\hat{X}^n$; the third terminal, the \emph{Receiver}, observes the side-information sequence $Y^n=(Y_1, \ldots, Y_n)\in \set{Y}^n$.  Under the null hypothesis 
\begin{align}
\mathcal{H}=0\colon \quad (X^n,Y^n)\sim \text{i.i.d.}\; P_{XY}, \label{eqn:null}
\end{align}
whereas under the alternative hypothesis 
\begin{align}
\mathcal{H}=1\colon \quad (X^n,Y^n)\sim \text{i.i.d.}\; Q_{XY},\label{eqn:alt}
\end{align}
for two given pmfs $P_{XY}$ and $Q_{XY}$. 

The privacy mechanism is described by the conditional pmf $P_{\hat{X}^n|X^n}$ which maps each sequence $X^n\in\mathcal{X}^n$ to a sequence $\hat{X}^n\in\hat{\mathcal{X}}^n$. For any $(\hat{x}^n,x^n,y^n)\in \hat{\mathcal{X}}^n\times \mathcal{X}^n\times \mathcal{Y}^n$, the joint distributions considering the privacy mechanism are given by 
\begin{align}
\!\! P_{\hat{X}XY}^n(\hat{x}^n,x^n,y^n) &\!\eqdef\! P_{\hat{X}^n|X^n}(\hat{x}^n|x^n)\cdot \prod_{i=1}^n  P_{XY}(x_i,y_i),\\
\!\! Q_{\hat{X}XY}^n(\hat{x}^n,x^n,y^n) &\!\eqdef\! P_{\hat{X}^n|X^n}(\hat{x}^n|x^n)\cdot\prod_{i=1}^n   Q_{XY}(x_i,y_i).
\end{align}

A \emph{memoryless/local} privacy mechanism is defined by a conditional pmf $P_{\hat{X}|X}$ which stochastically and  independently maps each entry $X_i\in \set{X}$ of $X^n$ to a released  $\hat{X}_i \in \hat{\set{X}}$ to construct $\hat{X}^n$. Consequently, for the memoryless privacy mechanism, the conditional pmf $P_{\hat{X}^n|X^n}(\hat{x}^n|x^n)$  factorizes
 as follows:
\begin{IEEEeqnarray}{rCl}
&&P_{\hat{X}^n|X^n}(\hat{x}^n|x^n)=\prod_{i=1}^n P_{\hat{X}|X}(\hat{x}_i|x_i)=P_{\hat{X}|X}^n(\hat{x}^n|x^n),\nonumber\\*&&\hspace{4.3cm} \forall (\hat{x}^n,x^n)\in \hat{\set{X}}^n\times \set{X}^n.\end{IEEEeqnarray}

There is a noise-free bit pipe of rate $R$ from the transmitter to the receiver. Upon observing $\hat{X}^n$, the transmitter computes the message $M=\phi^{(n)}(\hat{X}^n)$ using a possibly stochastic encoding function $\phi^{(n)}: \hat{\set{X}}^n\to \{0,\ldots,\lfloor 2^{nR} \rfloor\}$ and sends it over the bit pipe to the receiver.  

The goal of the receiver is to produce a guess of $\mathcal{H}$ using a decoding function $g^{(n)}:\mathcal{Y}^n\times \{0,...,\lfloor 2^{nR}\rfloor\}\to \{0,1\}$  based on the observation $Y^n$ and the received message $M$. Thus the estimate of the hypothesis is   $\hat{\mathcal{H}}=g^{(n)}(Y^n,M)$.

This induces a partition of the sample space $\hat{\set{X}}^n\times \set{X}^n\times \set{Y}^n$ into an acceptance region $\set{A}_n$ defined as follows:
\begin{align}
&\mathcal{A}_n \eqdef  \left\{(\hat{x}^n,x^n,y^n)\colon  g^{(n)}(y^n,\phi^{(n)}(\hat{x}^n))=0 \right\},
\end{align}
and a rejection region denoted by $\mathcal{A}^c_n$.

\begin{definition}\label{def} For any $\epsilon \in [0,1)$ and for a given rate-privacy pair $(R,L)\in \Reals_+^2$, we say that a type-II exponent $\theta\in \Reals_+$ is $(\epsilon,R,L)$-achievable 
	if there exists a sequence of functions and conditional pmfs $(\phi^{(n)},g^{(n)},P_{\hat{X}^n|X^n})$, such that the corresponding sequences of type-I  and type-II error probabilities at the receiver are respectively defined as
	\begin{align}\alpha_{n}\eqdef P_{\hat{X}XY}^n(\set{A}_n^c)\quad\mbox{and}\quad\beta_{n}\eqdef Q_{\hat{X}XY}^n(\set{A}_n),\end{align}
	and they satisfy 
	\begin{align}
	\limsup_{n\to\infty}\;	\alpha_{n}&\leq \epsilon\quad\text{and}\quad 	\liminf_{n\to\infty}\;\frac{1}{n}\log\frac{1}{\beta_{n}}\geq \theta.
	\end{align}
	Furthermore, the \emph{privacy measure}
\begin{equation}
T_n \eqdef \frac{1}{n}I(X^n;\hat{X}^n),
\end{equation}
satisfies 
	\begin{align}
	\limsup_{n\to\infty}\; T_n \leq L.
	\end{align}
	The {\em optimal exponent} $\theta_\epsilon^*(R,L)$ is the supremum of all $(\epsilon,R,L)$-achievable $\theta\in \Reals_+$.
\end{definition}

\section{General Hypothesis Testing} \label{sec:general}

\subsection{Achievable Error Exponent}
The  following presents an achievable error exponent for the proposed setup.
\begin{theorem}\label{achievability_general} 
	For a given $\epsilon \in [0,1)$ and a rate-privacy pair $(R,L)\in \Reals_+^2$, the optimal type-II error exponent $\theta^*_{\epsilon}(R,L)$ for the multiterminal hypothesis testing setup under the privacy constraint $L$ and the rate constraint $R$ satisfies
	\begin{IEEEeqnarray}{rCl}
		\theta^*_{\epsilon}(R,L)\ge \hspace{-0.1cm}\max_{\substack{P_{U|\hat{X}},P_{\hat{X}|X}:\\R\geq I(U;\hat{X})\\L\geq I(X;\hat{X})}}\;\min_{\substack{\tilde{P}_{U\hat{X}XY}\in \\\set{P}_{U\hat{X}XY}}}  D(\tilde{P}_{U\hat{X}XY}\|P_{U|\hat{X}}P_{\hat{X}|X}Q_{XY}\!),\nonumber\\\label{ach_errorex}
	\end{IEEEeqnarray}
	where the set $\set{P}_{U\hat{X}XY} $ is defined as
\begin{align}
	\set{P}_{U\hat{X}XY} &\stackrel{\Delta}{=} \left\{  \tilde{P}_{U\hat{X}XY} \,\, \left|\,\, \parbox[c]{.8 in}{$\tilde{P}_{X} = P_{X}$, \vspace{0.04 in}\\ $\tilde{P}_{UY} = P_{UY} $, \vspace{0.04 in} \\ $\tilde{P}_{U\hat{X}}=  P_{U\hat{X}}$}  \right. \right\}.
\end{align}	
	Given $P_{U|\hat{X}}$ and $P_{\hat{X}|X}$, the mutual informations in \eqref{ach_errorex} are calculated according to the following joint distribution:
	\begin{align}
	P_{U\hat{X}XY} &\eqdef P_{U|\hat{X}}\cdot P_{\hat{X}|X}\cdot P_{XY}.\label{eq:joint_dist}
	\end{align}
\end{theorem}
\begin{IEEEproof} The coding scheme is given in the following section. For the analysis, see Appendix~\ref{sec_ach}. 
\end{IEEEproof}

\subsection{Coding Scheme}\label{sec:coding-testing} 

In this section, we propose a coding scheme for Theorem~\ref{achievability_general}, under  fixed rate and privacy constraints $(R,L)\in \Reals_+^2$. Fix the joint distribution $P_{U\hat{X}XY}$ as in \eqref{eq:joint_dist}. Let $P_U(u)$ be the marginal distribution of $U\in\mathcal{U}$ defined as
\begin{align}
P_U(u) \eqdef \sum_{\hat{x}\in\hat{\set{X}}} P_{U|\hat{X}}(u|\hat{x})\sum_{x\in\mathcal{X}}P_{\hat{X}X}(\hat{x},x).\end{align} Fix positive $\mu>0$ and $\zeta>0$, an arbitrary blocklength $n$ and two conditional pmfs $P_{\hat{X}|X}$ and $P_{U|\hat{X}}$ over finite auxiliary alphabets $\hat{\mathcal{X}}$ and $\mathcal{U}$. Fix also the rate and privacy leakage level as 
\begin{align}
R  = I(U;\hat{X})+\mu, \quad\mbox{and}\quad
L  = I(\hat{X};X)+\zeta. \label{ratecons}
\end{align}

\underline{\textit{Codebook Generation}}: Randomly and independently generate a codebook
\begin{align}\mathcal{C}_U \eqdef \left\{ U^n(m): m\in\{0,\ldots, \lfloor 2^{nR}\rfloor\} \right\},\label{codebook}\end{align}
by drawing $U^n(m)$ in an i.i.d. manner according to $P_U$. The codebook is shown to all terminals.

\underline{\textit{Observer}}: Upon observing $x^n$, it checks whether ${x^n\in\mathcal{T}_{\mu/4}^n(P_X)}.$ If successful, it outputs the sequence $\hat{x}^n$ where its $i$-th component $\hat{x}_{i}$ is generated based on $x_i$, according to $P_{\hat{X}|X}(\hat{x}_i|x_i)$. If the typicality check is not successful, the observer then outputs $0^n$ which is an all-zero sequence of length $n$, where $\hat{x}^n= 0^n$. 

\underline{\textit{Transmitter}}: Upon observing $\hat{x}^n$, if $\hat{x}^n\neq 0^n$, the transmitter finds an index $m$ such that ${\big(u^n(m),\hat{x}^n\big)\in \mathcal{T}_{\mu/2}^n(P_{U\hat{X}})}.$ 
If successful, it sends the index $m$ over the noiseless link to the receiver. Otherwise, if the typicality check is not successful or $\hat{x}^n= 0^n$, it sends $m=0$. 

\underline{\textit{Receiver}}: Upon observing $y^n$ and receiving the index $m$, if $m=0$, the receiver declares $\hat{\mathcal{H}}=1$. If $m\neq0$, it checks whether ${\big(  u^n(m),y^n\big)\in \mathcal{T}_{\mu}^n(P_{UY})}.$
If the test is successful, the receiver declares $\hat{\mathcal{H}}=0$; otherwise, it sets $\hat{\mathcal{H}}=1$.

\begin{remark} In the above scheme, the sequence $\hat{X}^n$ is chosen to be an $n$-length zero-sequence when the observer finds that $X^n$ is not typical according to $P_X$. Thus, the privacy mechanism is not memoryless and the sequence $\hat{X}^n$ is not i.i.d. A detailed analysis  in Appendix~\ref{sec_ach} shows  that the privacy criterion is not larger than $L$ as the blocklength $n\to\infty$.
	\end{remark}

\subsection{Discussion}
In the following, we discuss some special cases. First, suppose that $R=0$.  As it is shown in the following corollary, Theorem~\ref{achievability_general} recovers Han's result  \cite{AhlswedeCsiszar} for distributed hypothesis testing with zero-rate communication.

\begin{corollary}[Theorem 5 in \cite{Han}]\label{thm3}  Suppose that $Q_{XY}>0$. For all $\epsilon \in [0,1)$, the optimal error exponent of the zero-rate communication for any privacy mechanism (including non-memoryless mechanisms) is given by the following:
	\begin{align}
	\theta_\epsilon^* (0,L)  =  \min_{\substack{\tilde{P}_{XY}:\\\tilde{P}_{X}=P_X\\\tilde{P}_Y=P_Y}} D(\tilde{P}_{XY}\|Q_{XY}\!).	 \label{eqn:nu}	\end{align}		
\end{corollary}
\begin{IEEEproof}  The proof of achievability follows by Theorem~\ref{achievability_general}, in which $\hat{X}$ is arbitrary and the auxiliary $U=\emptyset$ due to the zero-rate constraint. The proof of the strong converse  follows along the same lines as \cite{ShalabyPapamarcou}.\end{IEEEproof}
\begin{remark} Consider the case of $R>0$ and $L=0$ where $\hat{X}$ is independent of $X$. Using Theorem~\ref{achievability_general}, the optimal error exponent is lower bounded as follows:
	\begin{align}
	\theta_\epsilon^* (R,0)  \geq  \min_{\substack{\tilde{P}_{XY}:\\\tilde{P}_{X}=P_X\\\tilde{P}_Y=P_Y}} D(\tilde{P}_{XY}\|Q_{XY}\!).	 	\end{align}
	However, the above error exponent is not necessarily optimal since the communication-rate is positive. Comparing this special case with the one in Corollary~\ref{thm3} shows that the proposed model does not, in general, admit symmetry between the rate and privacy constraints.  However, we will see from some specific examples in the following that the roles of $R$ and $L$ are symmetric.
	\end{remark}

Now, suppose that $L$ is so large such that $L>H(X)$. The following corollary shows that Theorem~\ref{achievability_general} recovers Han's result in \cite{Han} for distributed hypothesis testing over a rate-$R$ communication link.
\begin{corollary}[Theorem 2 in \cite{Han}] Assuming $L>H(X)$, the optimal error exponent is lower bounded as the following:
	\begin{equation}
	\theta_{\epsilon}^*(R,L)\ge \max_{\substack{P_{U|X}:\\R\geq I(U;X)}}\;\;\; \min_{\substack{\tilde{P}_{UXY}:\\\tilde{P}_{UX}=P_{UX}\\\tilde{P}_{UY}=P_{UY}}} D(\tilde{P}_{UXY}\| P_{U|X}Q_{XY}).
	\end{equation}
\end{corollary}
\begin{IEEEproof} The proof follows from Theorem~\ref{achievability_general} by specializing to $\hat{X}=X$.
	\end{IEEEproof}

The above two special cases reveal a trade-off between the privacy criterion and the achievable error exponent when the communication rate is positive, i.e., $R>0$. An increase in $L$ results in a larger achievable error exponent. This observation is further illustrated by an example in Section~\ref{ex:binary} to follow.

\section{Hypothesis Testing Against Independence with A Memoryless Privacy Mechanism}\label{sec:positive-rate}
In this section, we consider testing against independence where the joint pmf under $\mathcal{H}=1$ factorizes as follows:
\begin{align}
Q_{XY} = P_X\cdot P_Y.
\end{align}
The privacy mechanism is assumed to be memoryless here. 

\subsection{Optimal Error Exponent}\label{sec:TAI}
The following theorem, which includes a strong converse, states the optimal error exponent for this special case.
\begin{theorem}\label{TAI_thm}
For any $(R,L)\in\Reals_+^2$, define 
\begin{equation}
\theta_\epsilon^*(R,L) = \;\;\max_{\substack{P_{U|\hat{X}},P_{\hat{X}|X}:\\\;R\geq I(U;\hat{X})\\\; L\geq I(X;\hat{X})}}\;\; I(U;Y).\label{th0}
\end{equation}
Then, for any   $\epsilon \in [0,1)$ and any $(R,L)\in\Reals_+^2$, the optimal error exponent for testing against independence when using a memoryless privacy mechanism is given by~\eqref{th0}, where it suffices to choose $|\mathcal{U}|\leq |\hat{\mathcal{X}}|+1$ and $|\hat{\mathcal{X}}|\leq |\mathcal{X}|$ according to Caratheodory's theorem~\cite[Theorem~15.3.5]{cover}.
\end{theorem}

%
\begin{IEEEproof}
The coding scheme is given in the following section. For the rest of proof,	see Appendix~\ref{conv_proof}. \end{IEEEproof}
\subsection{Coding Scheme}\label{ind:scheme}
In this section, we propose a coding scheme for Theorem~\ref{TAI_thm}.
Fix the joint distribution as in \eqref{eq:joint_dist}, and the rate and privacy constraints as in \eqref{ratecons}. Generate the codebook $\mathcal{C}_U$ as in \eqref{codebook}. 

\underline{\textit{Observer}}: Upon observing $x^n$,  it outputs the sequence $\hat{x}^n$ in which the $i$-th component $\hat{x}_{i}$ is generated based on $x_i$, according to~$P_{\hat{X}|X}(\hat{x}_i|x_i)$.

\underline{\textit{Transmitter}}: It finds an index $m$ such that ${\big(u^n(m),\hat{x}^n\big)\in \mathcal{T}_{\mu/2}^n(P_{U\hat{X}}).}$
If successful, it sends the index $m$ over the noiseless link to the receiver. Otherwise, it sends $m=0$. 

\underline{\textit{Receiver}}: Upon observing $y^n$ and receiving the index $m$, if $m=0$, the receiver declares $\hat{\mathcal{H}}=1$. If $m\neq0$, it checks whether $\big(  u^n(m),y^n\big)\in \mathcal{T}_{\mu}^n(P_{UY}).$ 
If the test is successful, the receiver declares $\hat{\mathcal{H}}=0$; otherwise, it sets $\hat{\mathcal{H}}=1$.

\begin{remark} In the above scheme, the sequence $\hat{X}^n$ is i.i.d.\ since it is generated based on the memoryless mechanism~$P_{\hat{X}|X}$.
	\end{remark}

When the  communication rate is positive, there exists a trade-off between the optimal error exponent and the privacy criterion. The following example elucidates this trade-off.

\subsection{Binary Example} \label{ex:binary}	
In this section, we study hypothesis testing against independence for a binary example. Suppose that under both hypotheses, we have $X\sim \text{Bern}(\frac{1}{2})$. Under the null hypothesis, \begin{align}\mathcal{H}=0\colon \qquad Y = X\oplus N,\qquad N\sim \text{Bern}(q)\end{align} for some $0\leq q\leq 1$, where $N$ is independent of $X$. Under the alternative hypothesis 
\begin{equation}
\mathcal{H}=1\colon \qquad Y\sim \text{Bern}\Big(\frac{1}{2}\Big),
\end{equation}
 where $Y$ is independent of $X$. The cardinality constraint shows that it suffices to choose $|\hat{\mathcal{X}}|=2$.  Due to symmetry of the source $X$ on its alphabet, without loss of optimality, we can choose $P_{\hat{X}|X}$  to be a binary symmetric channel (BSC). The argument follows since the error exponent depends on  ${X}$ through the conditional pmf $P_{U|\hat{X}}$ thanks to the Markov chain $U\markov \hat{X}\markov X$. The random variable ${\hat{X}}$ is determined by $P_{\hat{X}|X}$ through the privacy constraint $L\geq I(X;\hat{X})$.
This constraint remains unchanged by choosing $P_{\hat{X}|X}(1|0)=P_{\hat{X}|X}(0|1)$ and $P_{\hat{X}|X}(0|0)=P_{\hat{X}|X}(1|1)$ due to symmetry of the source $X$. 
	
	The cardinality bound on the auxiliary random variable $U$ is $|\mathcal{U}|\leq 3$. The following proposition states that  it is also optimal to choose $P_{U|\hat{X}}$ to be a BSC.

	\begin{proposition}\label{bin-ex-prop} The optimal error exponent of the proposed binary setup is given by the following:
			\begin{align}
		\theta_\epsilon^*(R,L)
		&= 1-h_{\text{b}}\left(q\star h_{\text{b}}^{-1}(1-L)\star h_{\text{b}}^{-1}(1-R)\right).\label{actual-final}
		\end{align}	
		\end{proposition}
	\begin{IEEEproof} For the proof of achievability, choose the following auxiliary random variables:
			\begin{align}\hat{X}&=X\oplus \hat{Z},\qquad \hat{Z}\sim \text{Bern}(p_1)\\ U&= \hat{X}\oplus Z,\qquad Z\sim \text{Bern}(p_2),\label{aux1}\end{align}
		for some $0\leq p_1,p_2\leq 1$ where $\hat{Z}$ and $Z$ are  independent of $X$ and $(X,\hat{X})$, respectively. The optimal error exponent of Theorem~\ref{TAI_thm} reduces to the following:
			\begin{align}
		\theta_\epsilon^*(R,L)&=\max_{\substack{0\leq p_1,p_2\leq 1:\\\;R\geq 1-h_{\text{b}}(p_2)\;\\L\geq 1-h_{\text{b}}(p_1)}} 1-h_{\text{b}}(q\star p_1\star p_2),
		\end{align}
		which can be simplified to \eqref{actual-final}.
		 For the proof of the converse, see Appendix~\ref{bin-ex:conv}.  
		\end{IEEEproof}

	\begin{figure}[t]
		\centerline{\includegraphics[scale=0.45]{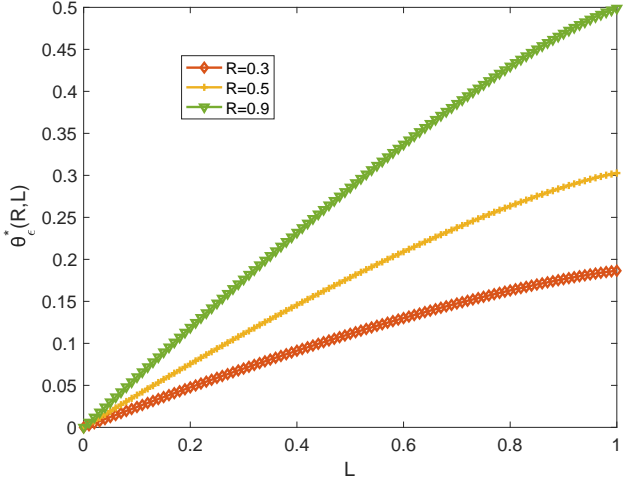}}
		\vspace*{-4mm}
		\caption{$\theta^*_\epsilon(R,L)$ versus $L$ for $q=0.1$ and various values of $R$.  \label{fig:exponent-privacy}}
		\vspace*{-3mm}
	\end{figure}
Fig.~\ref{fig:exponent-privacy} illustrates the error exponent versus the privacy parameter $L$ for  a fixed  rate $R$. There is clearly a trade-off between $\theta_\epsilon^*(R,L)$ and $L$. For a less stringent  privacy requirement  (large $L$), the error exponent $\theta_\epsilon^*(R,L)$ increases.

\subsection{Euclidean Approximation}\label{sec:Euc} 

In this section, we propose   Euclidean approximations~\cite{Borade,huang} for the optimal error exponent of testing against independence scenario (Theorem~\ref{TAI_thm}) when $R\approx 0$ and $L\approx 0$. Consider the optimal error exponent as follows:
\begin{align}
\theta_\epsilon^*(R,L) = \max_{\substack{P_{U|\hat{X}},P_{\hat{X}|X}:\\\;R\geq I(U;\hat{X})\; \\L\geq I(X;\hat{X})}}\;\; I(U;Y).\label{eq:57}
\end{align}
Let $\mathbf{W}$, of dimension $|\set{Y}|\times|\set{X}|$, denote the transition matrix $P_{Y|X}$, which is itself 
 induced by $P_{X}$ and the joint distribution $P_{XY}$. Now, consider the rate constraint as follows:
\begin{align}
I(U;\hat{X})&=\sum_{u\in \set{U}} P_{U}(u) D\big(P_{\hat{X}|U}(\cdot|u)\| P_{\hat{X}}\big)\le R.\label{eq:rate-142}
\end{align}
Assuming $R\approx 0$, we let $P_{\hat{X}|U}(\cdot|u)$ be a local perturbation from $P_{\hat{X}}(\cdot)$, where we have \begin{equation}
P_{\hat{X}|U}(\cdot|u)=P_{\hat{X}}(\cdot)+ \psi_u(\cdot),\label{approx1}
\end{equation} 
for a perturbation $\psi_u(\cdot)$ satisfying
\begin{align}
\sum_{\hat{x}\in\hat{\mathcal{X}}} \psi_u(\hat{x})=0,\label{dist-approx1}
\end{align}
in order to preserve the row stochasticity of $P_{\hat{X}|U}$. Using a $\chi^2$-approximation~\cite{Borade}, we can write:
\begin{align}
D \big(P_{\hat{X}|U}(\cdot|u)\| P_{\hat{X}}\big)&\approx  \frac12\cdot \log e\cdot\left\|\mathbf{k}_u\right\|^2,
\end{align}
where 
$\mathbf{k}_u$ denotes the  length-$|\hat{\mathcal{X}}|$ column vector of weighted perturbations whose $\hat{x}$-th component is defined as:
\begin{align}
k_u(\hat{x})&\eqdef \frac{1}{\sqrt{P_{\hat{X} }(\hat{x})    }} \cdot \psi_u(\hat{x}),\quad \forall \hat{x}\in \hat{\set{X}}.
\end{align}
%
Using the above definition, the rate constraint in \eqref{eq:rate-142} can be written as:
\begin{align}
\sum_{u\in\set{U}} P_{U}(u) \left\|\mathbf{k}_u\right\|^2 &\le \frac{2R}{\log e}.
\end{align}

Similarly, consider the privacy constraint as the following:
\begin{align}
I(X;\hat{X})
&=\sum_{\hat{x}\in\hat{\mathcal{X}}} P_{\hat{X}}(\hat{x}) D\big(P_{X|\hat{X}}(\cdot|\hat{x})\| P_{X}\big)\le L.\label{privacy_constraint}
\end{align}
Assuming $L\approx 0$, we let $P_{X|\hat{X}}(\cdot|\hat{x})$ be a local perturbation from $P_{X}(\cdot)$ where 
\begin{align}
P_{X|\hat{X}}(\cdot|\hat{x})=P_X(\cdot)+ \phi_{\hat{x}}(\cdot),\label{approx2}
\end{align}
for a perturbation $\phi_{\hat{x}}(\cdot)$ that satifies:
\begin{align}
\sum_{x\in\mathcal{X}}\phi_{\hat{x}}(x)=0.\label{dist-approx2}
\end{align}
Again, using a $\chi^2$-approximation, we obtain the following:
\begin{align}
D\big(P_{X|\hat{X}}(\cdot|\hat{x})\| P_{X}\big)\approx \frac{1}{2} \; \log e\; \left\|\mathbf{k}_{\hat{x}}\right\|^2,
\end{align}
where 
$\mathbf{k}_{\hat{x}}$ is a length-$|\mathcal{X}|$ column vector and its $x$-th component is defined as follows:
\begin{align}
k_{\hat{x}}(x)&\eqdef \frac{1}{\sqrt{P_{X }(x)    }} \cdot \phi_{\hat{x}}(x),\quad \forall x\in \set{X}.
\end{align}
%
Thus, the privacy constraint in \eqref{privacy_constraint} can be written as:
\begin{align}
\sum_{\hat{x}\in
	\hat{\set{X}}} P_{\hat{X}}(\hat{x}) \left\|\mathbf{k}_{\hat{x}}\right\|^2 &\le \frac{2L}{\log e}.
\end{align}
For any $x\in\mathcal{X}$ and $u\in\mathcal{U}$, we define the following:
\begin{align}
\Lambda_u(x)&\eqdef \sum_{\hat{x}\in\hat{\mathcal{X}}} \psi_u(\hat{x})\; \phi_{\hat{x}}(x)\\&=\sqrt{P_X(x)}\; \sum_{\hat{x}\in\hat{\mathcal{X}}} \sqrt{P_{\hat{X}}(\hat{x})}\; k_u(\hat{x})\; k_{\hat{x}}(x),\label{eta-def}
\end{align}
and the corresponding length-$|\set{X}|$ column vector $\mathbf{\Lambda}_u$ defined as follows:
\begin{align}
 \mathbf{\Lambda}_u = \left[\sqrt{P_X}\right] \mathbf{K}_{\hat{X}}\left[ \sqrt{P_{\hat{X}}}\right]\mathbf{k}_u,\label{eta-vec-def}
\end{align}
where $\left[\sqrt{P_X}\right]$ denotes a diagonal $|\mathcal{X}|\times |\mathcal{X}|$-matrix, so that its $(x,x)$-th element ($x\in\mathcal{X}$) is $\sqrt{P_X(x)}$, and $\left[ \sqrt{P_{\hat{X}}}\right]$ is  defined similarly. Moreover, $\mathbf{K}_{\hat{X}}$ refers to the $|\set{X}|\times|\hat{\set{X}}|$-matrix defined as follows:
\begin{align}
\mathbf{K}_{\hat{X}} \eqdef \left[ \begin{array}{cccccc} \mathbf{k}_1 & \mathbf{k}_2 & \ldots & \mathbf{k}_{\hat{x}} & \ldots & \mathbf{k}_{|\hat{\set{X}}|}\end{array} \right].
\end{align}
Let $\left[ \sqrt{P_Y} \right]^{-1}$ be the inverse of diagonal $|\mathcal{Y}|\times |\mathcal{Y}|$-matrix $\left[ \sqrt{P_Y} \right]$.
As shown in Appendix~\ref{Euc}, the optimization problem in \eqref{eq:57} can be written as follows:
\begin{align}
&\max_{\{\mathbf{k}_u\}_{u\in\set{U}},\mathbf{K}_{\hat{X}}}\qquad \frac{1}{2} \; \log e\; \Bigg[\sum_{u\in \set{U}} P_U(u) \cdot  \nonumber\\&\hspace{1.2cm}\left\|\left[ \sqrt{P_Y} \right]^{-1} \mathbf{W} \left[\sqrt{P_X}\right] \mathbf{K}_{\hat{X}}\left[ \sqrt{P_{\hat{X}}}\right]\mathbf{k}_u\right\|^2\Bigg]\label{opt-final}\\
&\text{subject to:} \quad\sum_{u\in\set{U}} P_{U}(u) \left\|\mathbf{k}_u\right\|^2 \le \frac{2R}{\log e},\\&\hspace{1.9cm} \sum_{\hat{x}\in\hat{\set{X}}} P_{\hat{X}}(\hat{x}) \left\|\mathbf{k}_{\hat{x}}\right\|^2 \le \frac{2L}{\log e}.\label{opt-final2}
\end{align}

The following example specializes the above approximation to the binary case.

\begin{example} Consider the binary setup of Example \ref{ex:binary} and the choice of auxiliary random variables in \eqref{aux1}. Since the privacy mechanism is assumed to be a BSC,  we have
	\begin{align}
	\mathbf{P}_X = \left[\frac{1}{2}\;\;\; \frac{1}{2}\right]^T,\qquad \mathbf{P}_{\hat{X}}=\left[\frac{1}{2}\;\;\; \frac{1}{2}\right]^T,
	\end{align}
	Now, we consider the vectors $\mathbf{k}_{u=0}$ and $\mathbf{k}_{u=1}$ defined as 
	\begin{align}
	\mathbf{k}_{u=0} &= \begin{bmatrix}\sqrt{2} \xi_1 & -\sqrt{2} \xi_1  \end{bmatrix}^T,\\
	\mathbf{k}_{u=1} &= \begin{bmatrix}-\sqrt{2} \xi_1&  \sqrt{2}\xi_1  \end{bmatrix}^T.
	\end{align}
	for some positive $\xi_1$.
	This yields the following:
	\begin{align}
	\mathbf{P}_{\hat{X}|U=0} &= \mathbf{P}_{\hat{X}}+\left[\xi_1\;\; -\xi_1\right]^T,\\\qquad \mathbf{P}_{\hat{X}|U=1} &= \mathbf{P}_{\hat{X}}+\left[-\xi_1\;\;\; \xi_1\right]^T
	\end{align}
	We also choose the vectors $\mathbf{k}_{\hat{x}=0}$ and $\mathbf{k}_{\hat{x}=1}$ as follows:
	\begin{align}
	\mathbf{k}_{\hat{x}=0} &= \begin{bmatrix}\sqrt{2} \xi_2 & -\sqrt{2} \xi_2\cdot \end{bmatrix}^T,\\\qquad
	\mathbf{k}_{\hat{x}=1} &= \begin{bmatrix}-\sqrt{2} \xi_2  & \sqrt{2} \xi_2 \end{bmatrix}^T,
	\end{align}
	which results in 
	\begin{align}
	\mathbf{P}_{X|\hat{X}=0} &= \mathbf{P}_{X}+\left[\xi_2\;\; -\xi_2\right]^T,\\\qquad \mathbf{P}_{X|\hat{X}=1} &= \mathbf{P}_{X}+\left[-\xi_2\;\;\; \xi_2\right]^T.
	\end{align}
	Notice that the matrix $\mathbf{W}$ is given by
	\begin{align}
	\mathbf{W} = \left[ \begin{array}{cc}  1-q & q \\ q & 1-q \end{array} \right].
	\end{align}
	Thus, the optimization problem in \eqref{opt-final} and \eqref{opt-final2} reduces to the following:
	\begin{align}
	&\max_{\xi_1,\xi_2}\qquad  8\;\log e\;  (1-2 q)^2   \;   |\xi_1|^2\; |\xi_2|^2\\
	&\text{subject  to:}   \qquad 4\; |\xi_1|^2 \le \frac{2R}{\log e} \quad \text{and}\quad 
	4\; |\xi_2|^2 \le \frac{2L}{\log e}.
	\end{align}
	Solving the above optimization yields
	\begin{align}
	\theta_\epsilon^*(R\approx 0,L\approx 0)&\approx 
\frac{2}{\log e}\; (1-2 q)^2   \;   R\; L.
	\label{approx-final}
	\end{align}
	For some values of parameters, the approximation in \eqref{approx-final} is compared to the error exponent of \eqref{actual-final} in Fig.~\ref{fig:exponent-privacy-approx}.   We observe that when $R=L\approx 0$, the approximation turns out to be excellent.
	\begin{figure}[t]
		\centerline{\includegraphics[scale=0.3]{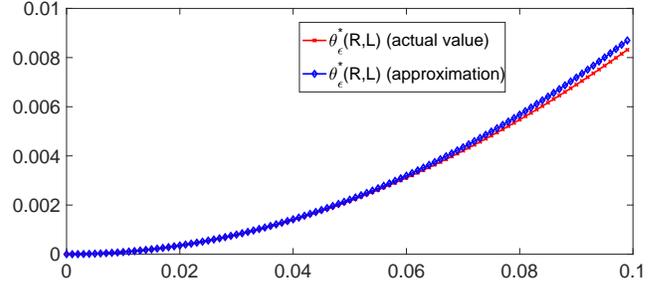}}
		\caption{$\theta_\epsilon^*(R\approx 0,L\approx 0)$ versus $L$ for $q=0.1$ and $R=L$. }
		\label{fig:exponent-privacy-approx}
	\end{figure}
	\end{example}

\begin{remark} The trade-off between the optimal error exponent and the privacy can again be verified from \eqref{approx-final} in the case of $L\approx 0$ and $R\approx 0$. As $L$ becomes larger (which corresponds to a less stringent privacy requirement), the error exponent also increases. For a fixed error exponent, a trade-off between $R$ and $L$ exists. An increase in $R$ results in a decrease of~$L$.
	\end{remark}

\subsection{Gaussian Setup}\label{sec:Gaussian}
In this section, we consider hypothesis testing against independence over a Gaussian example. 
Suppose that $X\sim \mathcal{N}(0,1)$ and under the null hypothesis $\mathcal{H}=0$, the sources $X$ and $Y$ are jointly Gaussian random variables distributed as $\mathcal{N}(0,\mathbf{G}_{XY})$, where $\mathbf{G}_{XY}$ is defined as the following:
	\begin{align}
	\mathbf{G}_{XY} \stackrel{\Delta}{=} \left[\begin{array}{ll} 1 & \rho \\ \rho & 1\end{array} \right],
	\end{align}
	for some $0\leq \rho \leq 1$.
	
	Under the alternative hypothesis $\mathcal{H}=1$, we assume that $X$ and $Y$ are independent Gaussian random variables, each distributed as $\mathcal{N}(0,1)$. Consider the privacy constraint as follows: 
	\begin{align}
	L &\geq I(X;\hat{X})
=h(X)-h(X|\hat{X}).\label{RHS}
	\end{align}
	For a Gaussian source $X$, the conditional entropy $h(X|\hat{X})$ is maximized for a jointly Gaussian $(X,\hat{X})$. This choice minimizes the RHS of \eqref{RHS}. Thus, without loss of optimality, we choose 
	\begin{align}
	X = \hat{X}+ Z, \qquad Z\sim \mathcal{N}\left(0,2^{-2L}\right),\label{aux-Gaussian}
	\end{align}
	where $Z$ is independent of $\hat{X}$. The following proposition states that it is optimal to choose $U$ jointly Gaussian with $(X,\hat{X},Y)$.
	\begin{proposition}\label{Gaussian-prop} The optimal error exponent of the proposed Gaussian setup is given by 
		\begin{align}
 \!\theta_\epsilon^*(R,L)\!	= \!	\frac{1}{2}\log \left( \frac{1}{1-\rho^2\cdot (1-2^{-2R})\cdot (1-2^{-2L})} \right).\label{ach-g-final}
		\end{align}
		\end{proposition}
	\begin{IEEEproof} For the proof of achievability, we choose $\hat{X}$ as in~\eqref{aux-Gaussian}. Also, let
		 \begin{align}
		 \hat{X}=U+\hat{Z},\qquad \hat{Z}\sim\mathcal{N}(0,\beta^2),\label{aux-Gaussian2}
		\end{align}
		for some $\beta^2\geq 0$, where $\hat{Z}$ is independent of $U$. For the details of the simplification and also the proof of converse, see Appendix~\ref{Ach-Gaus-proof}.
		\end{IEEEproof}
	
\begin{remark}	
	If $L=\infty$, the above proposition recovers the optimal error exponent of Rahman and Wagner~\cite[Corollary~7]{Wagner} for testing against independence of Gaussian sources over a noiseless link of rate~$R$.
	\end{remark}

\section{Summary and Discussion}\label{sec:conclusion}

In this paper, distributed hypothesis testing with privacy constraints is considered. A coding scheme is proposed where the sensor decides on one of hypotheses and generates the randomized data based on its decision. The transmitter describes the randomized data over a noiseless link to the receiver. The privacy mechanism in this scheme is non-memoryless.  The special case of testing against independence with a memoryless privacy mechanism is studied in detail. The optimal type-II error exponent of this case is established, together with  a strong converse. A binary example is proposed where the trade-off between the privacy criterion and the error exponent is reported.   Euclidean approximations are provided for the case in which the privacy level is high and  and the communication rate is vanishingly small. The optimal type-II error exponent of a Gaussian setup is also established. 

A future line of research is to study the second-order asymptotics of the proposed model. The second-order analysis of a distributed hypothesis testing without privacy constraints  and with zero-rate communication was studied in~\cite{Shun}. In all  our proposed extensions, the trade-off between the privacy and type-II error exponent is confirmed as an increase in the privacy criterion (a less stringent privacy requirement) yields a larger error exponent. The next step is to see whether the trade-off between privacy and error exponent affects  the second-order term.  

Another potential line for future research is to consider other metrics of privacy instead of the mutual information. A possible candidate is to use the maximal leakage \cite{Barthe-kopf,issa17, liao17} and to analyze the performance in tandem with  distributed hypothesis testing problem.

%
%
%
%
%

%

\appendices

\section{Proof of Theorem \ref{achievability_general}}
\label{sec_ach}

The analysis is based on the scheme of Section~\ref{sec:coding-testing}.

\underline{\textit{Error Probability Analysis:}}
We analyze type-I and type-II error probabilities averaged over all random codebooks. By standard arguments as in \cite[pp. 204]{cover}, it can be shown that there exists at least a codebook that satisfies the constraints on error probabilities. 

For the considered $\mu>0$ and the considered blocklength $n$, let $\mathcal{P}_{\mu}^n$ be the set of all joint types $\pi_{U\hat{X}XY}$ over $\set{U}^n\times \hat{\set{X}}^n\times \set{X}^n\times  \set{Y}^n$ which satisfy the following constraints:
\begin{align}
\big| \pi_{X}-P_{X}\big| &\leq \mu/4,\\
\big| \pi_{U\hat{X}}-P_{U\hat{X}} \big| &\leq \mu/2,\\
\big| \pi_{UY}-P_{UY} \big| &\leq \mu.
\end{align}
First, we analyze the type-I error probability. For the case of  $M\neq 0$, we define the following event:
\begin{align}
\mathcal{E}\eqdef \left\{(U^n(M),Y^n)\notin \mathcal{T}_{\mu}^n(P_{UY})\right\}.\label{Event}
\end{align}
Thus, type-I error probability can be upper bounded as follows:
\begin{align}
\alpha_n &\leq \Pr\left[\hat{X}^n=0^n\;\text{or}\;M=0\;\text{or}\;\mathcal{E} \big| \mathcal{H}=0\right]\\
&\leq \Pr\left[\hat{X}^n=0^n \big| \mathcal{H}=0\right]\nonumber\\&\qquad+\Pr\left[M=0\big|\hat{X}^n\neq 0^n,\mathcal{H}=0\right]\nonumber\\&\qquad+\Pr\left[\mathcal{E}\big|M\neq 0,\hat{X}^n\neq 0^n, \mathcal{H}=0\right]\\
&\le \epsilon/3+\Pr\left[M=0\big|\hat{X}^n\neq 0^n,\mathcal{H}=0\right]\nonumber\\&\qquad+\Pr\left[\mathcal{E}\big|M\neq 0,\hat{X}^n\neq 0^n ,\mathcal{H}=0\right] \label{eqn:aep}\\
&\le\epsilon/3+\epsilon/3+\Pr\left[\mathcal{E}\big|M\neq 0,\hat{X}^n\neq 0^n ,\mathcal{H}=0\right] \label{eqn:rate_con}\\
&\le\epsilon/3+\epsilon/3+\epsilon/3=\epsilon, \label{eqn:markov_lemma}
\end{align}
where \eqref{eqn:aep} follows from AEP~\cite[Theorem~3.1.1]{cover}; \eqref{eqn:rate_con} follows from the covering lemma~\cite[Lemma~3.3]{ElGamal} and the rate constraint \eqref{ratecons}, \eqref{eqn:markov_lemma} follows from Markov lemma~\cite[Lemma~12.1]{ElGamal}. In all justifications, $n$ is taken to be sufficiently large.

Next, we analyze the type-II error probability. 
The acceptance region at the receiver is  \begin{align}
&\set{A}_n^{\text{Rx}}=\bigcup_m   \Big\{\left(\hat{x}^n,x^n,y^n\right)\colon  \nonumber\\*
 &\qquad  \hat{x}^n\neq 0^n, \left(u^n(m),\hat{x}^n,x^n,y^n\right)\in\mathcal{T}_{\mu}^n(P_{U\hat{X}XY})\Big\}.
\end{align}
The set $\set{A}_n^{\text{Rx}}$  is contained within the following acceptance region $\bar{\mathcal{A}}_{n}$:
\begin{align}
&\bar{\set{A}}_{n}=\bigcup_m\bigg\{\left(\hat{x}^n,x^n,y^n\right)\colon \nonumber\\*
&\qquad\hat{x}^n\neq 0^n,\left(u^n(m),\hat{x}^n,x^n,y^n\right)\in\bigcup_{\pi\in \set{P}_\mu^n}\mathcal{T}^n(\pi)\bigg\}.\label{acceptance_region}
\end{align}
Let $\mathcal{F}_m \triangleq\{ \big(U^n(m),\hat{X}^n,X^n,Y^n\big)\!\in\!\set{P}_{\mu}^n\}$. 
Therefore, the average of type-II error probability over all codebooks is upper bounded as follows:
\begin{align}
\mathbb{E}_{\mathcal{C}}[\beta_n] & \leq  Q_{\hat{X}XY}^n\big(\bar{\set{A}}_{n}\big)\\
&\!\leq \sum_m\Pr\left[ \hat{X}^n\!\neq\! 0^n,\mathcal{F}_m \,\big|\,\set{H} = 1\right]\!\\
&\!\leq\sum_m\Pr\left[\mathcal{F}_m \, \big| \,\hat{X}^n\ne 0^n,\set{H} =1\right]\!\!\\
&\le 2^{nR}  \cdot  (n+1)^{ |\mathcal{U}|\cdot |\hat{\mathcal{X}} |\cdot |\mathcal{X}|\cdot |\mathcal{Y}|} \nonumber\\*
&\qquad\cdot  \max_{\pi_{U\hat{X}XY}\in\mathcal{P}_{\mu}^n}\;\;   2^{-n D(\pi_{U\hat{X}XY}\|P_UP_{\hat{X}|X}Q_{XY}) } \label{eqn:sanov}\\
&= (n+1)^{ |\mathcal{U}|\cdot |\hat{\mathcal{X}} |\cdot |\mathcal{X}|\cdot |\mathcal{Y}|}  \cdot 2^{-n\tilde{\theta}_{\mu}},
\end{align}
where
 \begin{equation}
 	\displaystyle{\tilde{\theta}_{\mu}\eqdef \min_{\pi_{U\hat{X}XY}\in\mathcal{P}_{\mu}^n} D(\pi_{U\hat{X}XY}\|P_UP_{\hat{X}|X}Q_{XY})-R},\end{equation}
and \eqref{eqn:sanov} follows from the upper bound of  Sanov's theorem~\cite[Theorem~11.4.1]{cover}. Hence, 
\begin{align}
\hspace*{-3mm}\tilde{\theta}_{\mu} &= \min_{\pi_{U\hat{X}XY}\in\mathcal{P}_{\mu}^n} D(\pi_{U\hat{X}XY}\|P_UP_{\hat{X}|X}Q_{XY})-R\\
&=\min_{\pi_{U\hat{X}XY}\in\mathcal{P}_{\mu}^n} D(\pi_{U\hat{X}XY}\|P_UP_{\hat{X}|X}Q_{XY}) \nonumber\\*
&\qquad\qquad\qquad-I(U;\hat{X})-\mu \label{eqn:rc}\\
&=\min_{\pi_{U\hat{X}XY}\in\mathcal{P}_{\mu}^n} D(\pi_{U\hat{X}XY}\|P_{U|\hat{X}}P_{\hat{X}|X}Q_{XY})+\delta(\mu),\label{errorex}
\end{align}
where $\delta(\mu)\to 0$ as $\mu\to 0$. Equality \eqref{eqn:rc} follows from the rate constraint in~\eqref{ratecons} and  \eqref{errorex} holds because $|\pi_{U\hat{X}}-P_{U\hat{X}}|<\mu/2$.

\underline{\textit{Privacy Analysis}}:  
We analyze the privacy when $\mathcal{H}=0$. A similar analysis holds for $\mathcal{H}=1$. Notice that $\hat{X}^n$ is not necessarily i.i.d.\ because according to the scheme in  Section~\ref{sec:coding-testing}, $\hat{X}^n$ is forced to be an all-zero sequence if the observer decides that $X^n$ is not typical. However, conditioned on the event that $X^n\in\mathcal{T}_{\mu}^n(P_X)$, the sequence $\hat{X}^n$ is i.i.d.\ according to the conditional pmf $P_{\hat{X}|X}$.
 The privacy measure $T_n$ satisfies
\begin{align}
nT_n &= I(X^n;\hat{X}^n)\\&=H(\hat{X}^n)-H(\hat{X}^n|X^n).
\end{align}

In the sequel, we provide a lower bound on $H(\hat{X}^n|X^n)$.
\begin{align}
H(\hat{X}^n|X^n)&=\sum_{x^n\in \set{X}^n} P_X^n(x^n) H(\hat{X}^n|X^n=x^n)\\
&\ge\!\!\!\!\!\! \sum_{x^n\in \set{T}_{\mu}^n(P_X)}\!\!\!\!\!\! P_X^n(x^n) H(\hat{X}^n|X^n=x^n)\label{eq:28}
\end{align}

For any $x^n \in \set{T}_{\mu}^n(P_X)$ and for $\mu'>\mu$, it holds that
\begin{align}
\lefteqn{H(\hat{X}^n|X^n=x^n)}\nonumber\\&=-\sum_{\hat{x}^n\in\hat{\set{X}}^n}P_{\hat{X}|X}^n(\hat{x}^n|x^n) \log P_{\hat{X}|X}^n(\hat{x}^n|x^n)\\
&\ge  -\!\!\!\!\!\!\!\!\sum_{\hat{x}^n\in\set{T}_{\mu'}^n(P_{\hat{X}|X}(\cdot|x^n)) }\!\!\!\!\!\!\!\! P_{\hat{X}|X}^n(\hat{x}^n|x^n) \log P_{\hat{X}|X}^n(\hat{x}^n|x^n)\\
&\ge  -\!\!\!\!\! \sum_{\hat{x}^n\in\set{T}_{\mu'}^n(P_{\hat{X}|X}(\cdot|x^n)) }\!\!\!  P_{\hat{X}|X}^n(\hat{x}^n|x^n) \nonumber\\*
&\qquad\qquad \qquad\qquad\times  \log \big[2^{-n (1-\mu')H(\hat{X}|X)}\big] \label{eqn:aa}\\
&\ge n (1-\mu')^2 H(\hat{X}|X)\label{eq:29}
\end{align}
where \eqref{eqn:aa} is true because for any $\hat{x}^n\in \set{T}_{\mu'}^n(P_{\hat{X}|X}(\cdot|x^n))$, it holds that $P_{\hat{X}|X}^n(\hat{x}^n|x^n) \le 2^{-n (1-\mu')H(\hat{X}|X)}$, and \eqref{eq:29} follows because the conditional typicality lemma~\cite[Chapter~2]{ElGamal} implies that $P_{\hat{X}|X}^n(\set{T}_{\mu'}^n(P_{\hat{X}|X}(\cdot|x^n) |x^n) \ge 1-\mu'$ for $n$ sufficiently large. 

Combining \eqref{eq:28} and \eqref{eq:29}, we obtain 
\begin{align}
H(\hat{X}^n|X^n)&\ge n (1-\mu')^2 H(\hat{X}|X)  \!\! \sum_{x^n\in \set{T}_{\mu}^n(P_X)}\!\!\!\!\!\! P_X^n(x^n) \\
&\ge n (1-\mu')^2 (1-\mu) H(\hat{X}|X), \label{eqn:cc} 
\end{align} 
where \eqref{eqn:cc} follows because the AEP~\cite[Theorem~3.1.1]{cover} implies that $P_X^n(\set{T}_{\mu}^n(P_X))\ge 1-\mu$ for $n$ sufficiently large.

Hence, we have
\begin{align}	
I(X^n;\hat{X}^n)&=H(\hat{X}^n)-H(\hat{X}^n|X^n)\\
&\le n H(\hat{X})-H(\hat{X}^n|X^n)\\
&\le n H(\hat{X}) - n (1-\mu'') H(\hat{X}|X)\\
&=nI(X;\hat{X})+ n \mu'' H(\hat{X}|X)\\
&\le n L + n \mu'' H(\hat{X}|X)\\
&\le nL+n\mu'' \cdot \log|\hat{\mathcal{X}}|\\
&=nL+n\zeta,
\end{align}
where $\mu''\eqdef 1-(1-\mu')^2(1-\mu) \ge 0$, and $\zeta\eqdef\mu''\cdot \log |\hat{\mathcal{X}}|$.

Letting  $n\to\infty$ and then letting $\mu,\mu'\to 0$, we obtain $\tilde{\theta}_{\mu}\to \theta$ and $\limsup_{n\to n} T_n\le L$, with $\theta$ given by the RHS of~\eqref{ach_errorex}. This establishes the proof of Theorem~\ref{achievability_general}.

\section{Proof of Theorem \ref{TAI_thm}}\label{conv_proof}
\underline{\textit{Achievability}}: The analysis is based on the scheme of Section~\ref{ind:scheme}. It follows similar steps as in \cite{AhlswedeCsiszar}. Recall the  definition of the  event $\mathcal{E}$ in~\eqref{Event}. Consider the type-I error probability as follows:
\begin{align}
\alpha_n &\leq \Pr\left[M=0\;\text{or}\;\mathcal{E} \big| \mathcal{H}=0\right]\\
&\leq \Pr\left[M=0\big|\mathcal{H}=0\right]+\Pr\left[\mathcal{E}\big|M\neq 0, \mathcal{H}=0\right]\\
&\le \epsilon/2+\epsilon/2\\&=\epsilon,\label{eqn:cover}
\end{align}
where \eqref{eqn:cover} follows from covering lemma \cite[Lemma~3.3]{ElGamal} and  the  rate constraint in \eqref{ratecons}, and also the Markov lemma~\cite[Lemma~12.1]{ElGamal}. Now, consider the type-II error probability as follows:
\begin{align}
\beta_n &= \Pr[\hat{\mathcal{H}}=0|\mathcal{H}=1]\\
&=\Pr[\hat{\mathcal{H}}=0,M\neq 0|\mathcal{H}=1]\\
&\leq \Pr[\hat{\mathcal{H}}=0|\mathcal{H}=1,M\neq 0]\\
&= \Pr[\hat{\mathcal{H}}=0|\mathcal{H}=1,M=1],
\end{align}
where the last equality follows from the symmetry of the code construction. Now, the average of type-II error probability over all codebooks satisfies:
\begin{align}
\mathbb{E}_{\mathcal{C}}\left[ \beta_n\right] \leq 2^{-n[I(U;Y)-\delta(\mu)]},
\end{align}
where $\delta(\mu)$ is a function that tends to zero as $\mu\to 0$. The privacy analysis is straightforward since the privacy mechanism is memoryless whence we have 
\begin{align}
\frac{1}{n}I(X^n;\hat{X}^n)=I(X;\hat{X})=L+\zeta,
\end{align}
where the last equality follows from the privacy constraint in \eqref{ratecons}. This concludes the proof of achievability.

\underline{\textit{Converse}}:
Now, we prove the strong converse. It involves an extension of the $\eta$-image characterization technique~\cite{CsiszarKorner,TianChen}.   For a given $P_{XY}$ define $V^n(y^n|x^n)\eqdef P_{Y|X}^n(y^n|x^n)$ for all $x^n\in\mathcal{X}^n$ and $y^n\in\mathcal{Y}^n$. A set $B \subseteq \set{Y}^n$ is an $\eta$-image of the set $A \subseteq  \set{X}^n$ over the channel $V^n$ if 
\begin{equation}
V^n\left(B|x^n\right)\ge \eta, \quad \forall x^n\in A.
\end{equation}
Let $\set{B}(A,\eta)$ denote the collection of all $\eta$-images of $A$ and define 
\begin{align}
\kappa_{V^n}(A,Q_{XY},\eta)\eqdef \frac{\min_{B\in \set{B}(A,\eta)} Q_{XY}^n(A\times B)}{P_{X}^n(A)}.
\end{align}
 This quantity is a generalization of the minimum cardinality of the $\eta$-images in \cite{CsiszarKorner} and is closely related to the minimum type-II error probability associated with the set $A$.

For the testing against independence setup, $Q_{XY}=P_X \cdot P_Y$, and thus
\begin{align}
\frac{Q_{XY}^n(A\times B)}{P_{X}^n(A)}=\frac{P_{X}^n(A)P_Y^n(B)}{P_{X}^n(A)}=P_Y^n(B),
\end{align}
and $\kappa_{V^n}(A,Q_{XY},\eta)$ is simply written as $\kappa_{V^n}(A,\eta)$ and is given by
\begin{align}
\kappa_{V^n}(A,\eta)\eqdef \min_{B\in \set{B}(A,\eta)} P_{Y}^n( B).
\end{align}

The proof of the upper bound on the error exponent in Theorem~\ref{TAI_thm} relies on the following lemma.
\begin{lemma}[Lemma 3 in \cite{TianChen}]\label{lem1}
	For any set $A\subseteq \set{X}^n$, consider a distribution $P_A^{(n)}$ over $A$ and let $P_A^{(n)}V^n$ be its corresponding output distribution induced by the channel $V^n$, i.e., \begin{equation}
	P_A^{(n)} V^n (y^n)\eqdef \sum_{x^n\in A } P_A^{(n)}(x^n) V^n\left(y^n|x^n\right).\end{equation}
	Then, for every $\delta'>0$, $0<\eta<1$, we have 
	\begin{align}
	\kappa_{V^n}(A,\eta) \ge  2^{-D(P_A^{(n)} V^n\| P_Y^n)-n\delta'}  
	\end{align} 
	for sufficiently large $n$.	
\end{lemma}

For any encoding function $\phi^{(n)}$ and any memoryless privacy mechanism $P_{\hat{X}|X}^n$ inducing an acceptance region $\mathcal{A}_n \subseteq \hat{\set{X}}^n\times \set{X}^n \times \set{Y}^n $, let $\tau_n$ denote the cardinality of codebook and define the following sets:
\begin{align}
C_i&\stackrel{\Delta}{=} \left\{\hat{x}^n\in\hat{\mathcal{X}}^n\colon \phi^{(n)}(\hat{x}^n)=i \right\},\\
D_i &\stackrel{\Delta}{=}  \left\{y^n\in\mathcal{Y}^n\colon g^{(n)}(y^n,i)=0 \right\},\quad 1\leq i\leq \tau_n.
\end{align}

The acceptance region can be written as follows:
\begin{align}
\mathcal{A}_n = \bigcup_{i=1}^{\tau_n} \left(C_i\times \mathcal{X}^n\times D_i\right),
\end{align}
where $C_i\cap C_j = \phi$ for all $i\neq j$. Define the set $\set{B}_n(\eta)$ as follows:
\begin{align}
&\set{B}_n(\eta)\eqdef \big\{(\hat{x}^n,x^n) \colon V^n\left(D_{\phi^{(n)}(\hat{x}^n)}| x^n\right)\ge \eta  \big\}.
\end{align}
Let $\mathcal{B}_n^x(\eta)$ be the projection of the above set onto $\mathcal{X}^n$, i.e.,
\begin{align}
\mathcal{B}_n^x(\eta)\eqdef \left\{ x^n\colon V^n\left(D_{\phi^{(n)}(\hat{x}^n)}| x^n\right)\ge \eta\;\text{for some}\; \hat{x}^n \right\}
\end{align}
Fix $\epsilon \in [0,1)$ and assume that the type-I error probability is upper-bounded as  
\begin{align}
\alpha_n &=P_{\hat{X}X Y}^n \left( \mathcal{A}_{n}^c\right)\leq \epsilon,
\end{align}
which we can write equivalently as
\begin{align}
1-\epsilon &  \le 	P_{\hat{X}X Y}^n \left(\set{A}_{n}\right)\\
&=\sum_{(\hat{x}^n,x^n)\in \set{B}_n(\eta) }
P_{\hat{X}X}^n(\hat{x}^n,x^n) V^n\left(D_{\phi^{(n)}(\hat{x}^n)}|x^n\right)\nonumber\\&+\sum_{(\hat{x}^n,x^n)\in \set{B}_n^c(\eta) }
P_{\hat{X}X}^n(\hat{x}^n,x^n) V^n\left(D_{\phi^{(n)}(\hat{x}^n)}|x^n\right)\\
&\le  P_{\hat{X}X}^n\left(\set{B}_n(\eta)\right)+ \eta   \left(1-P_{\hat{X}X}^n(\set{B}_n(\eta))\right),\label{eq:72}
\end{align}
where the first term is because $V^n\left(D_{\phi^{(n)}(\hat{x}^n)}|x^n\right)\le 1$; and the second term is because for any $(\hat{x}^n,x^n)\in \set{B}_n^c(\eta)$, we have $V^n\left(D_{\phi^{(n)}(\hat{x}^n)}|x^n\right)<\eta$. 

In what follows, let $\eta=\frac{1-\epsilon}{2}$. Inequality \eqref{eq:72} implies 
\begin{align}
P_{\hat{X}X}^n(\set{B}_n(\eta)) \ge \frac{1-\epsilon}{1+\epsilon}.
\end{align}
%
%
Let $\mu_n=n^{-1/3}$. For the typical set $\set{T}_{\mu_n}^n(P_{\hat{X}X})$, we have
\begin{align}
P_{\hat{X}X}^n\big(\set{T}_{\mu_n}^n(P_{\hat{X}X})\big)\geq 1-\frac{|\set{X}|\cdot |\hat{\set{X}}|}{4\mu_n^2 n}.
\end{align}

Hence,
\begin{align}
&P_{\hat{X}X}^n\!\left(\set{T}_{\mu_n}^n(P_{\hat{X}X})\cap\set{B}_n(\eta)\right)\nonumber\\
&\hspace{1cm}\ge P_{\hat{X}X}^n\left(\set{T}_{\mu_n}^n(P_{\hat{X}X})\right)+ P_{\hat{X}X}^n\left(\set{B}_n(\eta)\right)- 1\!\\
&\hspace{1cm}\ge \frac{1-\epsilon}{1+\epsilon} -\frac{|\set{X}|\cdot |\hat{\set{X}}|}{4\mu_n^2 n}.
\end{align}
For any $0<\delta< \frac{1-\epsilon}{1+\epsilon}$ and for sufficiently large $n$,
\begin{align}
P_{\hat{X}X}^n\big(\set{T}_{\mu_n}^n(P_{\hat{X}X})\cap \set{B}_n(\eta)\big)&\ge \delta.
\end{align}

We can also write $\set{T}_{\mu_n}^n(P_{\hat{X}X})$ as
\begin{align}
\mathcal{T}_{\mu_n}^n(P_{\hat{X}X}) = \bigcup_{\hat{P}_{\hat{X}X}:|\hat{P}_{\hat{X}X}-P_{\hat{X}X} |\leq \mu_n} \mathcal{T}^n(\hat{P}_{\hat{X}X}).
\end{align}
Combining the above equations, we get
\begin{align}
\sum_{\hat{P}_{\hat{X}X}:|\hat{P}_{\hat{X}X}-P_{\hat{X}X} |\leq \mu_n}\!\!\!\!	
P_{\hat{X}X}^n\left(\mathcal{T}^n(\hat{P}_{\hat{X}X})  \cap\set{B}_n(\eta)\right)\geq \delta.
\end{align}
Let $\tilde P_{\hat{X}X}$ denote the type which maximizes the $P_{\hat{X}X}^n$-probability  of the type class among all such types. As there exist at most $(n+1)^{|\hat{\set{X}}|\cdot |\set{X}|}$ possible types, it holds that
\begin{align}
P_{\hat{X}X}^n\big(\mathcal{T}^n(\tilde P_{\hat{X}X})  \cap\set{B}_n(\eta)\big)\ge \frac{\delta}{(n+1)^{|\hat{\set{X}}|\cdot |\set{X}|}}.\label{lb115}
\end{align}
Notice that the above inequality implies the following:
\begin{align}
P_{X}^n\big(\mathcal{T}^n(\tilde P_{X})  \cap\set{B}^x_n(\eta)\big)\ge \frac{\delta}{(n+1)^{|\hat{\set{X}}|\cdot |\set{X}|}},
\end{align} because $\Pr(A)\ge \Pr(A\cap B)$.
Define the sets $\Psi_n(\eta)\eqdef \mathcal{T}^n(\tilde P_{\hat{X}X})  \cap\set{B}_n(\eta)$ and $\Psi^x_n(\eta)\eqdef \mathcal{T}^n(\tilde P_{X})  \cap\set{B}_n^x(\eta)$. We can write the probability in \eqref{lb115} as 
\begin{align}
&P_{\hat{X}X}^n\big(\mathcal{T}^n(\tilde P_{\hat{X}X})  \cap\set{B}_n(\eta)\big) \nonumber\\
&=\sum_{ (\hat{x}^n,x^n) \in  \Psi_n(\eta)} P_{\hat{X}X}^n( \hat{x}^n,x^n) \\
&=  \sum_{ (\hat{x}^n,x^n) \in  \Psi_n(\eta)}2^{-n \big[D( \tilde{P}_{\hat{X}X} \| P_{\hat{X}X} ) + H_{\tilde{P}_{\hat{X}X}}( \hat{X},X)\big] }\\
&\le\sum_{ (\hat{x}^n,x^n) \in  \Psi_n(\eta)}2^{-n  [H( \hat{X},X ) -\delta_1 ] } \label{psi-ineq_pre}
\end{align}
where $\delta_1\to 0 $ as $n\to\infty$ due to the fact that $D( \tilde{P}_{\hat{X}X} \| P_{\hat{X}X} ) \ge 0$ and  $ | \tilde{P}_{\hat{X}X}  -  P_{\hat{X}X}|\le\mu_n$ so the entropies are also arbitrarily close.
It then follows from \eqref{lb115} and  \eqref{psi-ineq_pre} that 
\begin{align}
\frac{1}{n}\log |\Psi_n(\eta)| \geq H(\hat{X},X)-\delta_2,\label{psi-ineq}
\end{align} 
where $\delta_2\to 0$ as $\mu_n\to 0$. Similarly, we can show that
\begin{align}
\frac{1}{n}\log |\Psi^x_n(\eta)| \geq H(X)-\delta_3,\label{psi-x-ineq}
\end{align} 
where $\delta_3\to 0$ as $\mu_n\to 0$.

 The encoding function $\phi^{(n)}$ partitions the set $\Psi_n(\eta)$ into $\tau_n$ non-intersecting subsets $\{S_i\}_{i=1}^{\tau_n}$ such that $\phi^{(n)}(f^{(n)}(x^n))=i$ for any $x^n\in S_i$. Define the following distribution:
\begin{align}
\!\!P_{\hat{\munderbar{X}}^n\munderbar{X}^n}(\hat{x}^n,x^n)
&\eqdef\!\frac{P_{\hat{X}X}^n(\hat{x}^n,x^n) \cdot \mathds{1}\!\left\{(\hat{x}^n,x^n)\in \Psi_n(\eta) \right\} }{P_{\hat{X}X}^n(\Psi_n(\eta))}. 
\end{align}
Note that this distribution, denoted by $P_{\gamma}^{(n)}$, corresponds to a uniform distribution over the set $\Psi_n(\eta)$ because all the sequences in $\Psi_n(\eta)$ have the same type $\tilde P_{\hat{X}X}$, and as the probability is uniform on a type class under any i.i.d.\ measure. Hence, the resulting marginals $P_{\munderbar{\hat{X}}^n}$ and $P_{\munderbar{X}^n}$ are also uniform.

 Let $\munderbar{M}\eqdef\phi^{(n)} (\munderbar{\hat{X}}^n)$ and $\munderbar{Y}^n$ be connected with $\munderbar{X}^n$ by the channel $V^n=P_{Y|X}^n$. Also, let  $P_i^{(n)}V^n$ be the distribution of the random variable $\munderbar{Y}^n$ given $\munderbar{M}=i$.

The type-II error probability can be lower-bounded as:
\begin{align}
\beta_n&\ge\sum_{(\hat{x}^n,x^n)\in \Psi_n(\eta)}
P_{\hat{X}X}^n\left(\hat{x}^n,x^n\right)\cdot P_Y^n\left(D_{\phi^{(n)}(\hat{x}^n)}\right)\\
&=\sum_{i=1}^{\tau_n}
P_{\hat{X}X}^n(S_i)\cdot P_Y^n(D_i)\\
&  \ge  \sum_{i=1}^{\tau_n}
P_{\hat{X}X}^n(S_i)\cdot \kappa_{V^n}(S_i,\eta)\label{eqn:a}\\
&=P_{\hat{X}X}^n\left(\Psi_n(\eta)\right)\cdot  \sum_{i=1}^{\tau_n}
P_{\gamma}^{(n)}(S_i)\cdot \kappa_{V^n}(S_i,\eta)\\
&\ge 2^{-n\delta'} \cdot P_{\hat{X}X}^n\left(\Psi_n(\eta)\right) \nonumber\\*
&\qquad\cdot  \sum_{i=1}^{\tau_n}
P_{\gamma}^{(n)}(S_i)\cdot 2^{-D\left(P_i^{(n)} V^n\big\| P_Y^n\right)} \label{eqn:b}\\
&\ge 2^{-n\delta'}\cdot P_{\hat{X}X}^n\left(\Psi_n(\eta)\right) \nonumber\\*
&\qquad\cdot     2^{-   \sum_{i=1}^{\tau_n}
	P_{\gamma}^{(n)}(S_i)\cdot D\left(P_i^{(n)}V^n\big\| P_Y^n\right)}\label{eqn:c}\\
&\ge  \frac{2^{-n\delta'} \delta}{(n+1)^{|\hat{\set{X}}|\cdot|\set{X}|}}  \cdot 2^{-\sum_{i=1}^{\tau_n} P_{\gamma}^{(n)}(S_i)\cdot D\big(P_i^{(n)}V^n \big\| P_Y^n\big)},\!\! \label{eqn:d}
\end{align} 
where \eqref{eqn:a} follows from the definition of $\kappa_{V^n}(S_i,\eta)$, \eqref{eqn:b} follows because
Lemma~\ref{lem1} implies that for any distribution $P_i^{(n)}$ over the set $S_i$ it holds that
$\kappa_{V^n}(S_i,\eta)\ge 2^{-D\left(P_i^{(n)} V^n\| P_Y^n\right)-n\delta'}$, \eqref{eqn:c} follows because of the convexity of the function $t \mapsto 2^{t}$, and \eqref{eqn:d} follows by \eqref{lb115} and  the fact that $\Pr(A)\ge \Pr(A \cap B)$. Hence,
\begin{align}
\!\!\!-\frac1n \log \beta_n- \delta'' \le \frac1n \sum_{i=1}^{\tau_n} 
P_{\gamma}^{(n)}(S_i)\cdot D\left(P_i^{(n)} V^n\| P_Y^n\right),\!\! \label{eq:82}
\end{align}
where $\delta''\eqdef \delta'-\frac1n \log\frac{\delta}{(n+1)^{ |\hat{\set{X}}|\cdot|\set{X}|}}$.

 Considering the fact that $P_{\gamma}^{(n)}(S_i)=P_{\munderbar{M}}(i)$, the right-hand-side of \eqref{eq:82} can be upper-bounded as follows:
\begin{align}
&\frac1n \sum_{i=1}^{\tau_n} 
P_{\gamma}^{(n)}(S_i)\cdot D(P_i^{(n)}V^n\| P_Y^n)\nonumber\\
&=\frac1n \sum_{i=1}^{\tau_n}\sum_{y^n\in \set{Y}^n}  P_{\munderbar{M}\munderbar{Y}^n}(i,y^n)  \log\frac{P_{\munderbar{Y}^n|\munderbar{M}}(y^n|i)}{P_{Y}^n(y^n)}\\
&=-\frac1n  H(\munderbar{Y}^n|\munderbar{M}) -\frac1n \sum_{y^n\in \set{Y}^n}  P_{\munderbar{Y}^n}(y^n)  \log P_{Y}^n(y^n)\\
&=-\frac1n  H(\munderbar{Y}^n|\munderbar{M}) -\frac1n \sum_{y^n\in \set{Y}^n}  P_{\munderbar{Y}^n}(y^n)\sum_{t=1}^n  \log P_{Y}(y_t)\\
&=-\frac1n  H(\munderbar{Y}^n|\munderbar{M}) -\frac1n \sum_{t=1}^n  \sum_{\substack{y^n\in\mathcal{Y}^n}}  P_{\munderbar{Y}^n}(y^n)   \log P_{Y}(y_t)\\
&=-\frac1n  H(\munderbar{Y}^n|\munderbar{M}) -\frac1n \sum_{t=1}^n  \sum_{{y}_t\in \set{Y}}   P_{\munderbar{Y}_t}({y}_t)   \log P_{Y}(y_t)\\
&=-\frac1n  H(\munderbar{Y}^n|\munderbar{M}) +\frac1n \sum_{t=1}^n \left[H(\munderbar{Y}_t)+D(P_{\munderbar{Y}_t}\|P_{Y})\right]\\
&=\frac1n  \sum_{t=1}^n \left[H(\munderbar{Y}_t)- H(\munderbar{Y}_t|\munderbar{M}, \munderbar{Y}^{t-1})+D(P_{\munderbar{Y}_t}\|P_{Y})\right]\label{eqn:e}\\
&\le\frac{1}{n} \sum_{t=1}^n I(\munderbar{M},\munderbar{X}^{t-1},\hat{\munderbar{X}}^{t-1};\munderbar{Y}_t)+\frac{1}{n} \sum_{t=1}^n D(P_{\munderbar{Y}_t}\|P_{Y})\label{eqn:f}\\
&=\frac{1}{n} \sum_{t=1}^n I(\munderbar{U}_t;\munderbar{Y}_t)+\frac{1}{n} \sum_{t=1}^n D(P_{\munderbar{Y}_t}\|P_{Y})\label{eqn:g}\\
&=I(\munderbar{U};\munderbar{Y})+D(P_{\munderbar{Y}}\|P_{Y}).\label{eqn:h}
\end{align}
Here, \eqref{eqn:e}--\eqref{eqn:h} are justified in the following:
\begin{itemize}
	\item \eqref{eqn:e} follows by the chain rule;
	\item \eqref{eqn:f} follows  from the Markov chain $\munderbar{Y}^{t-1}\markov(\munderbar{M},\munderbar{X}^{t-1},\hat{\munderbar{X}}^{t-1})\markov \munderbar{Y}_t$;
	\item \eqref{eqn:g} follows from the definition \begin{align}\munderbar{U}_t\eqdef  (\munderbar{M},\munderbar{X}^{t-1},\hat{\munderbar{X}}^{t-1});\label{u-definition}\end{align}
	\item \eqref{eqn:h} follows by defining a time-sharing random variable $T$ over $\{1,\ldots,n \}$ and the following
	\begin{align} \munderbar{U}\stackrel{\Delta}{=} (\munderbar{U}_T,T),\;\; \munderbar{Y}\stackrel{\Delta}{=}\munderbar{Y}_T.\end{align}
	\end{itemize}
 This leads to the following upper-bound on the type-II error exponent: 
\begin{align}
-\frac1n \log \beta_n\le I(\munderbar{U};\munderbar{Y})+D(P_{\munderbar{Y}}\|P_{Y})+\delta''.
\end{align}
Next, the rate constraint satisfies the following:
\begin{align}
nR &\geq H(\munderbar{M})\\
&\geq I(\munderbar{M};\munderbar{X}^n,\hat{\munderbar{X}}^n)\\
&=H(\munderbar{X}^n,\hat{\munderbar{X}}^n)-H(\munderbar{X}^n,\hat{\munderbar{X}}^n|\munderbar{M})\\
&=\log \big|\Psi_n(\eta)\big|-H(\munderbar{X}^n,\hat{\munderbar{X}}^n|\munderbar{M})\label{eqn:i}\\
&\ge n(H(\hat{X},X)-\delta_2 )-H(\munderbar{X}^n,\hat{\munderbar{X}}^n|\munderbar{M}) \label{eqn:j}\\
&=nH(\hat{X},X)-\sum_{t=1}^n H(\munderbar{X}_t,\hat{\munderbar{X}}_t|\munderbar{X}^{t-1},\hat{\munderbar{X}}^{t-1},\munderbar{M}) \nonumber\\*
&\qquad\qquad-n\delta_2\\
& = nH(\hat{X},X)-\sum_{t=1}^n H(\munderbar{X}_t,\hat{\munderbar{X}}_t|\munderbar{U}_t)-n\delta_2 \label{eqn:k}\\
&= nH(\hat{X},X)-nH(\munderbar{X},\hat{\munderbar{X}}|\munderbar{U})-n\delta_2\label{eqn:l}
\end{align}
where \eqref{eqn:i} follows because the distribution $P_{\hat{\munderbar{X}}^n\munderbar{X}^n}$ is uniform over the set $\Psi_n(\eta)$; \eqref{eqn:j} follows from \eqref{psi-ineq}; \eqref{eqn:k} follows from the definition in \eqref{u-definition}; \eqref{eqn:l} follows by defining $\munderbar{X}\eqdef\munderbar{X}_T$ and $\hat{\munderbar{X}}\eqdef\hat{\munderbar{X}}_T$.

Finally, the privacy measure satisfies the following:
\begin{align}
n L &\geq  I(\munderbar{X}^n;\hat{\munderbar{X}}^n)\\
&= H(\munderbar{X}^n)-H(\munderbar{X}^n|\hat{\munderbar{X}}^n)\\
&= \log \big|\Psi^x_n(\eta)\big|-H(\munderbar{X}^n|\hat{\munderbar{X}}^n)\\
&\ge(H(X)-\delta_3)-H(\munderbar{X}^n|\hat{\munderbar{X}}^n)\label{eqn:m}\\
&=n(H(X)-\delta_3)-\sum_{t=1}^n H(\munderbar{X}_t|\munderbar{X}^{t-1},\hat{\munderbar{X}}^n)\\
&\geq n(H(X)-\delta_3)-\sum_{t=1}^n H(\munderbar{X}_t|\hat{\munderbar{X}}_t)\\
&= nH(X)-n H(\munderbar{X}|\hat{\munderbar{X}})-n\delta_3,\label{leakageweak}
\end{align}
where \eqref{eqn:m} follows from \eqref{psi-x-ineq} and  \eqref{leakageweak} follows by the usual time-sharing arguments.

 Since $\Psi_n(\eta)\subseteq \mathcal{T}^n(\tilde P_{\hat{X}X})$,
for any $x\in  \set{X}$ and $\hat{x}\in  \hat{\set{X}}$, 
\begin{align}
P_{\hat{\munderbar{X}}\munderbar{X}}(\hat{x},x)&=\frac1n \sum_{t=1}^n P_{\hat{\munderbar{X}}_t\munderbar{X}_t}(\hat{x},x)\\
&=\sum_{(\hat{x}^n,x^n)\in\Psi_n(\eta)} \frac{N\left(\hat{x},x|\hat{x}^n,x^n\right)}{n\cdot |\Psi_n(\eta)|}\\* &=  \tilde{P}_{\hat{X}X}(\hat{x},x). \label{eqn:bar_hat_eq}
\end{align}

Recall that $|\tilde{P}_{\hat{X}X}-P_{\hat{X}X} |\leq \mu_n$ with $\mu_n=n^{-1/3}$. Hence, from \eqref{eqn:bar_hat_eq}, it holds that $|P_{\hat{\munderbar{X}}\munderbar{X}}-P_{\hat{X}X} |\leq \mu_n$.
By the definitions of $\munderbar{\hat{X}}$, $\munderbar{{X}}$ and $\munderbar{Y}$, we can suppose $P_{Y|{X}}=P_{\munderbar{Y}|\munderbar{{X}}}=V$. 
The random variable $U$ is chosen over the same alphabet as $\munderbar{U}$ and such that $P_{U|\hat{X}}=P_{\munderbar{U}|\munderbar{\hat{X}}}$. 

Since $P_Y(y)>0$ for all $y\in\set{Y}$, letting $n\to \infty$ and $\mu_n\to 0$ and the uniform continuity of the involved information-theoretic quantities yields the following upper bound on the optimal error exponent:
\begin{IEEEeqnarray}{rCl}
	\theta_{\epsilon}^*(R,L) \leq I(U;Y),	
	\end{IEEEeqnarray}
subject to the rate constraint:
\begin{IEEEeqnarray}{rCl}
	R \geq I(U;\hat{X},X) \geq I(U;\hat{X}),\end{IEEEeqnarray}
	and the privacy constraint:
	\begin{IEEEeqnarray}{rCl}
	L &\geq & I(X;\hat{X}).
	\end{IEEEeqnarray}
This concludes the proof of converse.


\section{Proof of the Converse of Proposition \ref{bin-ex-prop}}\label{bin-ex:conv}
We simplify Theorem~\ref{TAI_thm} for the proposed binary setup. As discussed in Section~\ref{ex:binary}, from the fact that $|\hat{\mathcal{X}}|=2$ and the symmetry of the source $X$ on its alphabet, without loss of optimality, we can choose $P_{\hat{X}|X}$ to be a BSC. First, consider the rate constraint:
\begin{align}
	R &\geq  I(U;\hat{X})\\
	&= H(\hat{X})-H(\hat{X}|U)\\
	&=1-H(\hat{X}|U),
\end{align}
which can be equivalently written as the following:
\begin{align}
H(\hat{X}|U) &\geq 1-R.\label{pr-bin2}
\end{align}
Also, the privacy criterion can be simplified as follows:
\begin{align}
L &\geq I(\hat{X};X)\\
&=H(\hat{X})-H(\hat{X}|X)\\
&=1-H(\hat{X}|X)\\
&=1-H(\hat{Z}),
\end{align}
which can be equivalently written as
\begin{align}
H(\hat{Z}) \geq 1-L.\label{pr-bin}
\end{align}
Now, consider the error exponent $\theta$ as follows: 
\begin{align}
	\theta &\leq  I(U;Y)\\
	&= H(Y)-H(Y|U)\\
	&= H(Y)-H(X\oplus N|U)\\
	&= H(Y)-H(\hat{X}\oplus \hat{Z}\oplus N|U)\\
	&\le H(Y)-h_{\text{b}}\big(h_{\text{b}}^{-1}(H(\hat{X}|U))\star h_{\text{b}}^{-1}(1-L)\star q\big) \label{eqn:gerber}\\
	&\le H(Y)-h_{\text{b}}\big(h_{\text{b}}^{-1}(1-R)\star h_{\text{b}}^{-1}(1-L)\star q\big), \label{eqn:follows}
\end{align}
where \eqref{eqn:gerber} follows from Mrs.~Gerber's lemma~\cite[Theorem~1]{Wyner}  and the fact that $(\hat{Z},N)$ is independent of $U$ and also from~\eqref{pr-bin}; \eqref{eqn:follows} follows from \eqref{pr-bin2}. This concludes the proof of the proposition.

\section{Euclidean Approximation of Testing Agianst Independence}\label{Euc}
We analyze the Euclidean approximation with the parameters defined in Section~\ref{sec:Euc}.
Notice that since $U\markov \hat{X}\markov X\markov Y$ forms a Markov chain, it holds that, for any $u\in \set{U}$,
\begin{align}
\mathbf{P}_{Y|U=u}=\mathbf{W} \mathbf{P}_{X|U=u}.
\end{align}
Now, consider the following chain of equalities for any $x\in\mathcal{X}$:
\begin{align}
&P_{X|U}(x|u) \nonumber\\*
&=\sum_{\hat{x}\in\hat{\mathcal{X}}} P_{X\hat{X}|U}(x,\hat{x}|u)\\
&=\sum_{\hat{x}\in\hat{\mathcal{X}}} P_{\hat{X}|U}(\hat{x}|u)\; P_{X|\hat{X}, U}(x|\hat{x},u)\\
&=\sum_{\hat{x}\in\hat{\mathcal{X}}} P_{\hat{X}|U}(\hat{x}|u)\; P_{X|\hat{X}}(x|\hat{x}) \label{eqn:euc1}\\
&= \sum_{\hat{x}\in\hat{\mathcal{X}}} \left(P_{\hat{X}}(\hat{x})+ \psi_u(\hat{x})\right)\; \left(P_X(x)+ \phi_{\hat{x}}(x)\right)\label{eqn:euc2}\\
&=P_X(x)+ \sum_{\hat{x}\in\hat{\mathcal{X}}} \psi_u(\hat{x})\; \phi_{\hat{x}}(x)\nonumber\\&\quad+ \sum_{\hat{x}\in\hat{\mathcal{X}}} P_{\hat{X}}(\hat{x})\; \phi_{\hat{x}}(x)+ P_X(x)\; \sum_{\hat{x}\in\hat{\mathcal{X}}}\psi_u(\hat{x})\\
& = P_X(x)+ \sum_{\hat{x}\in\hat{\mathcal{X}}} \psi_u(\hat{x})\; \phi_{\hat{x}}(x),\label{eqn:euc3}
\end{align}
where \eqref{eqn:euc1}---\eqref{eqn:euc3} are justified in the following:
\begin{itemize}
	\item \eqref{eqn:euc1} follows from the Markov chain $U\markov \hat{X}\markov X$ where given $\hat{X}$, $U$ and $X$ are independent;
	\item \eqref{eqn:euc2}  follows from \eqref{approx1} and \eqref{approx2};
	\item \eqref{eqn:euc3}  follows from \eqref{dist-approx1} and also from \eqref{approx2} which yields the following:
	\begin{align}
	\sum_{\hat{x}\in\hat{\mathcal{X}}} P_{\hat{X}}(\hat{x})\cdot \phi_{\hat{x}}(x)=0.
	\end{align}
\end{itemize}
With the definition of $\Lambda_u(x)$ in \eqref{eta-def}, we can write
\begin{align}
P_{X|U}(x|u) &=P_X(x)+ \Lambda_u(x),\qquad \forall x\in\mathcal{X},u\in\mathcal{U}.
\end{align}
Thus, we get
\begin{align}
\mathbf{P}_{Y|U=u}&=\mathbf{W} \mathbf{P}_{X}+ \mathbf{W}  \mathbf{\Lambda}_u\\
&=\mathbf{P}_Y+\mathbf{W}  \mathbf{\Lambda}_u.\label{PYu}
\end{align}
Applying the  $\chi^2$-approximation and using \eqref{PYu}, we can rewrite $I(U;Y)$ as follows:
\begin{align}
I(U;Y)&\approx \frac{1}{2}\;\log e\;  \sum_{u\in \set{U}} P_U(u)\;  \left\| \left[ \sqrt{P_Y} \right]^{-1} \mathbf{W}  \mathbf{\Lambda}_u\right\|^2
\end{align}
The above approximation with the definition of the vector $\mathbf{\Lambda}_u$ in \eqref{eta-vec-def} yields the optimization problem in \eqref{opt-final}.

\section{Proof of Proposition \ref{Gaussian-prop}}\label{Ach-Gaus-proof}
\underline{\textit{Achievability}}:
We specialize the achievable scheme of Theorem~\ref{TAI_thm} to the proposed Gaussian setup. We choose the auxiliary random variables as in \eqref{aux-Gaussian} and \eqref{aux-Gaussian2}. Notice that from the Markov chain $U\markov \hat{X}\markov X\markov Y$ and also the Gaussian choice of $\hat{X}$ in \eqref{aux-Gaussian} which was discussed in Section~\ref{sec:Gaussian}, we can write $Y=\rho \hat{X}+F$ where $F\sim \mathcal{N}\left(0,1- \rho^2\cdot(1-2^{-2L})\right)$ is independent of $\hat{X}$.  These choices of auxiliary random variables lead to the following rate constraint:
\begin{align}
R\geq \frac{1}{2}\log \left( \frac{1-2^{-2L}}{\beta^2} \right),\label{achg1}
\end{align}
which can be equivalently written as:
\begin{align}
2^{-2R}\cdot \left(1-2^{-2L}\right)\leq \beta^2.\label{Gaussian-cons1}
\end{align}
The optimal error exponent is also lower bounded as follows
\begin{align}
\theta_{\epsilon}^*(R,L) \geq \frac{1}{2}\log \left( \frac{1}{1-\rho^2\cdot \left(1-2^{-2L}-\beta^2\right)}  \right).\label{achg3}
\end{align}
Combining \eqref{Gaussian-cons1} and \eqref{achg3} gives the lower bound on the error exponent in \eqref{ach-g-final}.

\underline{\textit{Converse}}: Consider the following upper bound on the optimal error exponent in Theorem~\ref{TAI_thm}:
\begin{align}
&\hspace{-0.1cm}\theta_{\epsilon}^*(R,L)\nonumber\\
&\leq  I(U;Y)\\
&=h(Y)-h(Y|U)\\
&= \frac{1}{2}\log \left(2\pi e \right)-h(Y|U)\\
&=\frac{1}{2}\log \left(2\pi e \right)-h\big(\rho \hat{X}+F\big|U\big)\\
&\le  \frac{1}{2}\log \left(2\pi e \right)-\frac{1}{2}\log \Big(  2^{2h\left( \rho \hat{X} |U \right)} \nonumber\\*
&\qquad\qquad+2\pi e \left(1-\rho^2\cdot (1-2^{-2L})\right) \Big)\label{eqn:epi}\\
&\leq  \frac{1}{2}\log \left(2\pi e \right)-\frac{1}{2}\log \Big(  \rho^2\; 2^{2h\left( \hat{X}|U \right)}\nonumber\\*
&\qquad\qquad +2\pi e \left(1-\rho^2\cdot(1-2^{-2L})\right)  \Big),\label{Gaussian-proof1}
	\end{align}
where \eqref{eqn:epi} follows from the entropy power inequality (EPI)~\cite[Chapter~2]{ElGamal}. Now, consider the rate constraint as follows:
\begin{align}
R&\geq I(\hat{X};U)\\
&=h(\hat{X})-h(\hat{X}|U)\\
&=\frac{1}{2}\log \left(2\pi e \left(1-2^{-2L}\right)\right)-h(\hat{X}|U),
\end{align}
which is equivalent to
\begin{align}
2^{2h(\hat{X}|U)}\geq 2\pi e\cdot 2^{-2R}\cdot \left(1-2^{-2L}\right).\label{Gaussian-proof2}
\end{align}
Considering \eqref{Gaussian-proof1} with \eqref{Gaussian-proof2} yields the following upper bound on the error exponent:
\begin{align}
 \theta_{\epsilon}^*(R,L)
	&\leq  \frac{1}{2}\log \left(2\pi e \right)-\frac{1}{2}\log \Big( 2\pi e \rho^2 2^{-2R}\left(1-2^{-2L} \right) \nonumber\\*
&\qquad	+2\pi e \left(1-\rho^2\; (1-2^{-2L})\right) \Big)\\
	&= \frac{1}{2}\log \left( \frac{1}{1-\rho^2\; (1-2^{-2R})\; (1-2^{-2L})} \right).
	\end{align}
This concludes the proof of the proposition.

\subsubsection*{Acknowledgements}
The authors would like to thank Mr.\ Lin Zhou (National University of Singapore) for helpful discussions during the preparation  of the manuscript.

\bibliographystyle{IEEEtran}
\bibliography{references}

\end{document}